\documentclass[sigconf,screen]{acmart}

\usepackage{amsmath,amssymb,amsfonts}
\usepackage{algorithmic}
\usepackage{graphicx}
\usepackage{textcomp}
\usepackage{xcolor}
\usepackage{csquotes}
\usepackage{multirow}
\usepackage{array}
\usepackage{todonotes}
\presetkeys{todonotes}{inline}{}
\usepackage[vlined,boxed]{algorithm2e}
\SetKwProg{Fn}{Function}{}{}
\usepackage{adjustbox}
\usepackage{autobreak}
\usepackage[normalem]{ulem}
\usepackage{arydshln}
\usepackage{mathtools}
\usepackage{booktabs}
\usepackage{ifsym}
\usepackage{fixltx2e}
\usepackage{tabularx}
\usepackage{subcaption}
\usepackage{MnSymbol}

\newcommand{\pglong}{property graph\xspace}
\newcommand{\pg}{PG\xspace}

\newcommand{\ingraph}{\mbox{ingraph}\xspace}

\newcommand{\iqd}{\mbox{\textsc{IncQuery-D}}\xspace}

\newcommand{\viatra}{\mbox{\textsc{Viatra}}\xspace}

\newcommand{\opencypher}{\mbox{openCypher}\xspace}

\newcommand{\eg}{e.g.\xspace}
\newcommand{\ie}{i.e.\xspace}

\definecolor{self}{HTML}{4daf4a}
\definecolor{read}{HTML}{377eb8}
\definecolor{prog}{HTML}{ff7f00}
\definecolor{todo}{HTML}{e41a1c}

\newcommand{\citeprog}[1]{\cite{#1}} 

\newcommand{\gra}{GRA\xspace}
\newcommand{\nra}{NRA\xspace}
\newcommand{\fra}{FRA\xspace}

\newcommand*{\mydots}{\ifmmode\mathellipsis\else.\kern-0.13em.\kern-0.13em.\fi} 

\newcommand{\vertexlabelfunction}{\mathit{lbl}}
\newcommand{\edgelabelfunction}{\mathit{typ}}

\newcommand{\vertexlabels}{L}
\newcommand{\edgelabels}{T}

\newcommand{\verticestoedges}{\mathit{st}}

\newcommand{\vertexproperties}{P_v}
\newcommand{\edgeproperties}{P_e}

\newcommand{\yes}{$\bigotimes$\xspace}
\newcommand{\maybe}{$\bigoslash$\xspace}
\newcommand{\no}{$\bigcirc$\xspace}

\newcommand{\nfii}{NF\textsuperscript{2}\xspace}

\newcommand{\yedscale}{0.33}

\newcommand{\examplespacing}{\vspace{0.6ex}}

\newcommand{\labelscell}{0.8cm}

\newcommand{\propstablewidth}{2.0cm}
\newcommand{\propscelli}{0.7cm}
\newcommand{\propscellii}{0.75cm}

\newcommand{\myvspace}{\vspace*{0.8ex}}

\newcommand{\metadata}{\scriptsize \sf}

\newcommand{\labelsi}[1]{
  \begin{tabular}{|p{\labelscell}|}
		\hline
		\metadata #1    \\ \hline
	\end{tabular}
}
\newcommand{\labelsii}[2]{
  \begin{tabular}{|p{\labelscell}|}
		\hline
		\metadata #1 \\ \hline
		\metadata #2 \\ \hline
	\end{tabular}
}

\newcommand{\propsi}[2]{
  \begin{minipage}{\propstablewidth}
  \myvspace
  \centering
  \begin{tabular}{|p{\propscelli}|p{\propscellii}|}
  	\hline
  	\scriptsize \sf #1 & #2 \\ \hline
  \end{tabular}
  \myvspace
  \end{minipage}
}

\newcommand{\propsiii}[6]{
  \begin{minipage}{\propstablewidth}
  \myvspace
  \centering
  \begin{tabular}{|p{\propscelli}|p{\propscellii}|}
  	\hline
  	\scriptsize \sf #1 & #2 \\ \hline
  	\scriptsize \sf #3 & #4 \\ \hline
  	\scriptsize \sf #5 & #6 \\ \hline
  \end{tabular}
  \myvspace
  \end{minipage}
}

\newcommand{\itemsi}[1]{
  [#1]
}
\newcommand{\itemsii}[2]{
  [#1, #2]
}

\newcommand{\vertexrow}[3]{
		$#1$ & #2 & \multicolumn{2}{c|}{#3}
}

\newcommand{\edgerow}[5]{
    #1 & $#2$ & $#3$ & \sf #4 & \multicolumn{2}{c|}{#5}
}

\newcolumntype{P}[1]{>{\centering\arraybackslash}p{#1}}

\expandafter\let\expandafter\pnewline\csname\string\ \endcsname

\newcommand{\grv}{\mathbb{V}}
\newcommand{\gre}{\mathbb{E}}

\newcommand{\fakeparagraph}[1]{\paragraph*{#1}}

\usepackage{ifxetex}
\usepackage{ifluatex}
\newif\ifxetexorluatex 
\ifnum 0\ifxetex 1\fi\ifluatex 1\fi>0
   \xetexorluatextrue
\fi

\ifxetexorluatex
  \usepackage{fontspec}
\else
  \usepackage[T1]{fontenc}
  \usepackage[utf8]{inputenc}
\fi

\usepackage{amsmath}
\usepackage{amssymb}
\usepackage{etoolbox}
\usepackage{xspace}
\newtoggle{textualoperators}

\usepackage{tikz}
\usepackage{forest}

\usetikzlibrary{arrows,shapes.arrows,automata}

\tikzset{every node/.style={draw}}

\usepackage{tensor}

\usepackage{fp}

\newcommand{\assign}{\rightarrow}

\newcommand{\asc}{\uparrow}
\newcommand{\desc}{\downarrow}

\newcommand{\tuple}[1]{\langle #1 \rangle}
\newcommand{\schematuple}[1]{\left\langle #1 \right\rangle}

\newcommand{\literal}[1]{\mathsf{#1}}
\newcommand{\atom}[1]{\mathsf{#1}}

\newcommand{\var}[1]{\mathit{#1}}

\newcommand{\tikzunit}{1.5ex}

\FPpow\triangleheight{3}{0.5}
\FPdiv\triangleheight{\triangleheight}{2}

\newtoggle{leftouterjoin}
\newtoggle{semijoin}
\newtoggle{antijoin}

\newcommand{\tikzjointemplate}[1]{%
	{
	#1
	\text{\,%
		\tikz[baseline, x=\tikzunit*\triangleheight, y=\tikzunit, join=round, cap=round]{
			\draw (-1,0) -- (0,0.5) -- (-1,1) -- cycle;
			\draw (+1,0) -- (0,0.5) -- (+1,1);
			\ifboolexpr{ not ( togl {semijoin} or togl {antijoin} ) }
			{
				\draw (+1,0) -- (+1,1);
			}{}
			\iftoggle{leftouterjoin}
			{
				\draw (-1,0) -- (-1.5,0);
				\draw (-1,1) -- (-1.5,1);
			}{}
			\iftoggle{antijoin}
			{
				\draw (-1.2,1.2) -- (+1.2,1.2);
			}{}
		}%
\,%
}}}

\newcommand{\joinsymbol}{%
	\tikzjointemplate{}%
}
\newcommand{\antijoinsymbol}{%
	\tikzjointemplate{\toggletrue{antijoin}}%
}
\newcommand{\leftouterjoinsymbol}{%
	\tikzjointemplate{\toggletrue{leftouterjoin}}%
}
\newcommand{\semijoinsymbol}{%
	\tikzjointemplate{\toggletrue{semijoin}}%
}

\newcommand{\vertexfree}{\medcircle}
\newcommand{\vertexmaybe}{\ocirc}
\newcommand{\vertexbound}{\odot}

\newcommand{\getverticesop}{\vertexfree}

\newcommand{\edgearrow}[1]{\mkern-2.8mu #1 \mkern-2.6mu}

\newcommand{\getedgesopdirected}  {{\getverticesop}{\edgearrow{\rightarrow}    }{\getverticesop}}
\newcommand{\getedgesopundirected}{{\getverticesop}{\edgearrow{\leftrightarrow}}{\getverticesop}}

\newcommand{\expandbothop}{{\vertexbound}{\edgearrow{\leftrightarrow}}{\vertexmaybe}}
\newcommand{\expandoutop} {{\vertexbound}{\edgearrow{\rightarrow}    }{\vertexmaybe}}
\newcommand{\expandinop}  {{\vertexbound}{\edgearrow{\leftarrow}     }{\vertexmaybe}}

\newcommand{\duplicateeliminationop}{\delta}
\newcommand{\sortop}{\tau}
\newcommand{\sortandtopop}{\topop\sortop}

\newcommand{\projectionop}{\pi}

\newcommand{\selectionop}{\sigma}
\newcommand{\groupingop}{\gamma}
\newcommand{\topop}{\lambda}
\newcommand{\unwindop}{\omega}

\newcommand{\joinop}{\joinsymbol}

\newcommand{\antijoinop}{\antijoinsymbol}
\newcommand{\semijoinop}{\semijoinsymbol}
\newcommand{\leftouterjoinop}{\leftouterjoinsymbol}
\newcommand{\thetaleftouterjoinop}{\leftouterjoinsymbol_\theta}
\newcommand{\unionop}{\cup}
\newcommand{\Unionop}{\bigcup}

\newcommand{\cartesianproductop}{\times}

\makeatletter
\newlength{\negph@wd}
\DeclareRobustCommand{\negphantom}[1]{%
  \ifmmode
    \mathpalette\negph@math{#1}%
  \else
    \negph@do{#1}%
  \fi
}
\newcommand{\negph@math}[2]{\negph@do{$\m@th#1#2$}}
\newcommand{\negph@do}[1]{%
  \settowidth{\negph@wd}{#1}%
  \hspace*{-\negph@wd}%
}

\newcommand{\getvertices}[2]{\getverticesop_{\var{#1}}^{\atom{#2}}}

\newcommand{\getedges}[7]{\tensor*[^{\atom{#2}}_{\var{#1}}]{{\vertexfree}{\edgearrow{#7[\var{#5}]{\atom{#6}}}}{\vertexfree}}{^{\atom{#4}}_{\var{#3}}}}
\newcommand{\getedgesdirected}[6]{\getedges{#1}{#2}{#3}{#4}{#5}{#6}{\xrightarrow}}
\newcommand{\getedgesundirected}[6]{\getedges{#1}{#2}{#3}{#4}{#5}{#6}{\xleftrightarrow}}

\newcommand{\dual}{\{\tuple{}\}}

\newcommand{\kleenestar}{\ast}

\newcommand{\expandedgevariable}[3]{
  {
	\atom{#1}
	\ifstrequal{#2}{1} 
	{
		\ifstrequal{#3}{1}
		{} 
		{\kleenestar_\atom{#2}^\atom{#3}} 
	} 
	{\kleenestar_\atom{#2}^\atom{#3}}
  }}

\newcommand{\expand}[8]{{}_{\var{#1}} {\vertexbound}{\edgearrow{#8[\var{#4}]{\expandedgevariable{#5}{#6}{#7}}}}{\vertexmaybe}_\var{#2}^\atom{#3}}

\newcommand{\expandboth}[7]{\expand{#1}{#2}{#3}{#4}{#5}{#6}{#7}{\xleftrightarrow}}
\newcommand{\expandout }[7]{\expand{#1}{#2}{#3}{#4}{#5}{#6}{#7}{\xrightarrow}    }
\newcommand{\expandin  }[7]{\expand{#1}{#2}{#3}{#4}{#5}{#6}{#7}{\xleftarrow}     }

\newcommand{\unwind}[1]{\unwindop_{\var{#1}}}
\newcommand{\unwindarrow}{\Rightarrow}

\newcommand{\duplicateelimination}{\duplicateeliminationop}
\newcommand{\sort}[1]{\sortop_{#1}}
\newcommand{\projection}[1]{\projectionop_{#1}}

\newcommand{\topp}[2]{\topop_{#1}^{#2}}
\newcommand{\sortandtop}[3]{\left\{ \sortop_{#1} \topop_{#2}^{#3} \right\}}

\newcommand{\selection}[1]{\selectionop_{#1}}

\newcommand{\grouping}[2]{\groupingop_{#1}^{#2}}

\newcommand{\join}{\joinop}
\newcommand{\transitivejoin}[2]{\overset{\ast_{#1}^{#2}}{\joinop}}
\newcommand{\antijoin}{\antijoinop}
\newcommand{\semijoin}{\semijoinop}
\newcommand{\leftouterjoin}{\leftouterjoinop}

\newcommand{\union}{\unionop}
\newcommand{\Union}{\Unionop}

\newcommand{\cartesianproduct}{\cartesianproduct}

\definecolor{red}{HTML}{d7191c}
\definecolor{lilac}{HTML}{984ea3}
\definecolor{red}{HTML}{d7191c}
\definecolor{blue}{HTML}{2b83ba}

\colorlet{nestedschemacolor}{red}
\colorlet{flatschemacolor}{blue}
\colorlet{nullarynodecolor}{lilac}

\newcommand{\operatortext}[1]{#1\xspace}

\newcommand{\getverticestext}{\operatortext{get-vertices}}
\newcommand{\getedgestext}{\operatortext{get-edges}}

\newcommand{\expandtext}{\operatortext{expand}}
\newcommand{\expandouttext}{\operatortext{expand-out}}
\newcommand{\expandintext}{\operatortext{expand-in}}
\newcommand{\expandbothtext}{\operatortext{expand-both}}

\newcommand{\transitiveexpandtext}{\operatortext{transitive expand}}
\newcommand{\transitiveexpandouttext}{\operatortext{transitive expand-out}}

\newcommand{\projectiontext}{\operatortext{projection}}

\newcommand{\selectiontext}{\operatortext{selection}}

\newcommand{\duplicateeliminationtext}{\operatortext{duplicate-elimination}}
\newcommand{\groupingtext}{\operatortext{grouping}}
\newcommand{\sorttext}{\operatortext{sort}}
\newcommand{\toptext}{\operatortext{top}}
\newcommand{\sortandtoptext}{\operatortext{sort-and-top}}
\newcommand{\unwindtext}{\operatortext{unwind}}

\newcommand{\antijointext}{\operatortext{antijoin}}
\newcommand{\semijointext}{\operatortext{semijoin}}

\newcommand{\transitivejointext}{\operatortext{transitive join}}
\newcommand{\leftouterjointext}{\operatortext{left outer join}}

\newcommand{\jointext}{\operatortext{natural join}}

\newcommand{\relnull}{\mathsf{NULL}}

\newcommand{\schema}{\mathrm{sch}}

\newcommand{\labelsLlong}{l_1, \ldots, l_\var{k}}
\newcommand{\labelsL}{L}

\newcommand{\labelsAlong}{l_{1,1}, \ldots, l_{1,\var{m}}}
\newcommand{\labelsA}{L1}

\newcommand{\labelsBlong}{l_{2,1}, \ldots, l_{2,\var{n}}}
\newcommand{\labelsB}{L2}

\newcommand{\typeslong}{t_1, \ldots, t_\var{o}}
\newcommand{\types}{T}

\newcommand{\alias}{/}
\usepackage{listings}

\usepackage{textcomp}
\definecolor{keyword}{HTML}{2771a3}
\definecolor{pattern}{HTML}{b53c2f}
\definecolor{string}{HTML}{be681c}
\definecolor{relation}{HTML}{7e4894}
\definecolor{variable}{HTML}{107762}
\definecolor{comment}{HTML}{8d9094}

\lstdefinelanguage{cypher}
{
	morekeywords={
		MATCH, OPTIONAL, WHERE, NOT, AND, OR, XOR, RETURN, DISTINCT, ORDER, BY, ASC, ASCENDING, DESC, DESCENDING, UNWIND, AS, UNION, WITH, ALL, CREATE, DELETE, DETACH, REMOVE, SET, MERGE, SET, SKIP, LIMIT, IN,
		INDEX, DROP, UNIQUE, CONSTRAINT, EXPLAIN, PROFILE, START, CASE,
		GROUP, HAVING,
	},
	sensitive=true,
	morecomment=[l]{//},
	morecomment=[s]{/*}{*/},
	morestring=[b]{"},
}

\newcommand{\mycdots}{\cdot\!\cdot\!\cdot}
\graphicspath{{./figures/}}

\setcopyright{none}
\acmYear{2018}
\copyrightyear{2018}

\begin{document}

\newcommand{\gc}{\mathit{gc}}

\newsavebox{\returnbox}
\begin{lrbox}{\returnbox}\lstinline+RETURN+\end{lrbox}
\newcommand{\lstreturn}{\usebox{\returnbox}}

\newsavebox{\withbox}
\begin{lrbox}{\withbox}\lstinline+WITH+\end{lrbox}
\newcommand{\lstwith}{\usebox{\withbox}}

\newsavebox{\matchbox}
\begin{lrbox}{\matchbox}\lstinline+MATCH+\end{lrbox}
\newcommand{\lstmatch}{\usebox{\matchbox}}

\newsavebox{\wherebox}
\begin{lrbox}{\wherebox}\lstinline+WHERE+\end{lrbox}
\newcommand{\lstwhere}{\usebox{\wherebox}}

\newsavebox{\distinctbox}
\begin{lrbox}{\distinctbox}\lstinline+DISTINCT+\end{lrbox}
\newcommand{\lstdistinct}{\usebox{\distinctbox}}

\newcommand{\lstreturnwith}{\lstreturn\ / \lstwith\ }

\SetKwFunction{InferRequiredProperties}{InferReq}
\SetKwFunction{ExtractProperties}{ExtractProperties}

\title[Reducing Property Graph Queries to Relational Algebra for IVM]{Reducing Property Graph Queries to Relational Algebra \\ for Incremental View Maintenance}

\newcommand{\bmemit}{Budapest University of Technology and Economics, \\ Department of Measurement and Information Systems}
\newcommand{\bmetmit}{Budapest University of Technology and Economics, \\ Dept. of Telecommunications and Media Informatics}

\newcommand{\mta}{MTA-BME Lendület Cyber-Physical Systems Res. Group}
\newcommand{\mcgill}{McGill University}

\settopmatter{authorsperrow=2}

\author{G\'abor Sz\'arnyas}
\orcid{0000-0001-8233-4431}
\affiliation{%
  \institution{\bmemit \\ \mta}
}
\email{szarnyas@mit.bme.hu}

\author{J\'ozsef Marton}
\orcid{0000-0003-4752-4234}
\affiliation{%
  \institution{\bmetmit}
}
\email{marton@db.bme.hu}

\author{J\'anos Maginecz}
\affiliation{%
  \bmemit
}

\author{D\'aniel Varr\'o}
\orcid{0000-0002-8790-252X}
\affiliation{%
  \institution{\bmemit, \mta, \mcgill}
}
\email{varro@mit.bme.hu}

\renewcommand{\shortauthors}{G. Sz\'arnyas et al.}

\begin{abstract}
The property graph data model of modern graph database systems is increasingly adapted for storing and processing heterogeneous datasets like networks. Many challenging applications with near real-time requirements -- e.g. financial fraud detection, recommendation systems, and on-the-fly validation -- can be captured with graph queries, which are evaluated repeatedly. To ensure quick response time for a changing data set, these applications would benefit from applying incremental view maintenance (IVM) techniques, which can perform continuous evaluation of queries and calculate the changes in the result set upon updates. However, currently, no graph databases provide support for incremental views. While IVM problems have been studied extensively over relational databases, views on property graph queries require operators outside the scope of standard relational algebra. Hence, tackling this problem requires the integration of numerous existing IVM techniques and possibly further extensions. In this paper, we present an approach to perform IVM on property graphs, using a nested relational algebraic representation for property graphs and graph operations.
Then we define a chain of transformations to reduce most property graph queries to flat relational algebra and use techniques from discrimination networks (used in rule-based expert systems) to evaluate them. We demonstrate the approach using our prototype tool, ingraph, which uses openCypher, an open graph query language specified as part of an industry initiative. However, several aspects of our approach can be generalised to other graph query languages such as G-CORE and PGQL.

\end{abstract}

\maketitle

\section{Introduction}
\label{sec:introduction}

\fakeparagraph{Concepts}
Graph processing problems are common in modern database systems, where the \emph{\pglong} (\pg) data model~\cite{DBLP:conf/edbt/HolschG16,DBLP:conf/adbis/MartonSV17,DBLP:conf/sigmod/AnglesABBFGLPPS18,DBLP:conf/sigmod/FrancisGGLLMPRS18,DBLP:conf/amw/Angles18} is gaining widespread 
adoption.
Property graphs extend labelled graphs with properties for both vertices and edges.
Compared to previous graph modelling approaches, such as the RDF data model (which treats properties as triples),
PGs allow users to store their graphs in a more compact and comprehensible representation.

\fakeparagraph{openCypher}
Due to the novelty of the \pg data model, no standard query language has emerged yet. The \emph{\opencypher initiative} aims to standardise the Cypher language~\cite{DBLP:conf/sigmod/FrancisGGLLMPRS18} of the Neo4j graph database.
The \opencypher language uses a SQL-like syntax
and combines graph pattern matching with relational operators (aggregations, joins, etc.).
In this paper, we target queries specified in the \opencypher language.

\fakeparagraph{Motivation}
In graph database applications, numerous use cases rely on complex queries and require low response time for repeated executions, including financial fraud detection, and recommendation engines.
In addition, graph databases are increasingly used in software engineering context as a semantic knowledge base for model validation~\cite{DBLP:conf/models/BergmannHRVBBO10,TrainBenchmarkSOSYM,DBLP:journals/scp/DanielSBTVGC17}, source code analysis~\cite{DBLP:conf/sigmod/HawesBC14}, etc.
While these scenarios could greatly benefit from \emph{incremental query evaluation}, currently no system provides incremental views with sufficient feature coverage for a practical PG query language such as \opencypher. Up to our best knowledge, the only existing incremental property graph query engine is Graphflow~\cite{DBLP:conf/sigmod/KankanamgeSMCS17},
which extends Cypher with triggers,
but lacks support for rich language features like negative/optional edges and transitive closures.

Incremental \emph{graph queries} were successfully used in the domain of model-driven engineering. For example, the incremental query engine of \viatra ensures quick model validation and transformation over in-memory graph models~\cite{DBLP:journals/scp/UjhelyiBHHIRSV15}. 

\fakeparagraph{Problem statement}
In relational database systems, \emph{incremental view maintenance} (IVM) techniques have been used for decades for repeated evaluation of a predefined query set on continuously changing data~\cite{DBLP:journals/ai/Forgy82,DBLP:conf/sigmod/BlakeleyLT86,DBLP:journals/tkde/MirankerL91,DBLP:conf/sigmod/GuptaMS93,DBLP:journals/debu/GuptaM95,DBLP:journals/tkde/Hanson96,Gupta:1999:MVT:310709,DBLP:journals/tkde/HansonBC02,DBLP:conf/models/SzarnyasIRHBV14,DBLP:journals/vldb/KochAKNNLS14,DBLP:journals/scp/UjhelyiBHHIRSV15}. However, these techniques typically build on assumptions that do not hold for property graph queries. In particular, \pg queries present numerous challenges:

\begin{enumerate}
\item
  \label{item:tackled-challenge-from}
  \label{challenge:schema-optional}
  \emph{Schema-optional data model.}
  Existing IVM techniques assume that the database schema is known a priori. While this is a realistic assumption for relational databases, the data model of most property graph systems is schema-free or schema-optional at best~\cite{DBLP:conf/sigmod/FrancisGGLLMPRS18}. Hence, to use IVM, users are required to manually define the schema of the graph, which is a tedious and error-prone process.

\item
  \emph{Nested data structures.}
  \label{challenge:nested}
  Most IVM techniques assume relational data model with 1NF relations.
  However, the property graph data model defines rich structures, including the properties on graph elements and paths. Collection types, such as sets, bags, lists, and maps are also allowed~\cite{DBLP:conf/sigmod/AnglesABBFGLPPS18,DBLP:conf/sigmod/FrancisGGLLMPRS18}. These can be represented as \nfii (non-first normal form) data structures, but their mapping to 1NF relations is a complex challenge.

\item
  \label{item:tackled-challenge-to}
  \label{challenge:meta}
  \emph{Mix of instance- and meta-level data.}
  Queries can not only access data fields from the instance graph (\eg ids, properties), but also metadata such as vertex labels and edge types~\cite{DBLP:conf/sigmod/FrancisGGLLMPRS18,DBLP:conf/grades/RestHKMC16}.

\item
  \label{item:naive-challenge-from}
  \label{challenge:nulls}
  \emph{Handling null values and outer joins.}
  Property graph queries allow null values and optional pattern matches, similarly to outer joins in relational databases. Most relational IVM works do not consider this challenge, except~\cite{DBLP:journals/sigmod/GriffinK98,DBLP:conf/icde/LarsonZ07}.

\item
  \emph{Complex aggregations.}
  \label{challenge:aggregation}
  PG queries allow complex aggregations, \eg
  aggregations on aggregations~\cite{DBLP:conf/sigmod/MumickQM97} and
  using non-distributive aggregation functions (\eg min, max, stdev) which are difficult to calculate incrementally~\cite{DBLP:conf/vldb/PalpanasSCP02}.

\item
  \label{item:naive-challenge-to}
  \label{challenge:reachability}
  \emph{Reachability queries.}
  Unbounded reachability queries on graphs with few connected components need to calculate large transitive closures, which makes them inherently expensive~\cite{DBLP:conf/gg/BergmannRSTV12}.
  Hence, the impact of the IVM on reachability is more limited compared to non-recursive queries
  and using space-time tradeoff techniques is more expensive: to improve execution time, one has to trade memory at an exponential rate.

\item
  \label{item:future-challenge-from}
  \label{challenge:transformations}
  \emph{Mix of queries and transformations.}
  Some property graph query languages (\eg \opencypher) allow combining update operations with queries. Most traditional IVM techniques do not consider this challenge, and omit related issues such as conflict set resolution. \emph{Discrimination networks} from rule-based expert systems are better suited to handle this issue~\cite{DBLP:journals/ai/Forgy82,DBLP:journals/tkde/MirankerL91,DBLP:journals/tkde/HansonBC02}.

\item
  \label{challenge:lists}
  \emph{List handling.}
  Property graph data sets and queries make use of lists both as a way to store collection of primitive values and to represent paths in the graphs. Order-preserving techniques have only been studied in the context of IVM on XQuery expressions~\cite{DBLP:conf/er/DimitrovaER03}, for trees but not for graphs.

\item
  \label{item:future-challenge-to}
  \label{challenge:skew}
  \emph{Skewed data distribution.}
  Subgraph matching is often implemented as a series of binary joins. Recent work revealed that binary (two-way) joins are inefficient on data sets with skewed distributions of certain edge types (displayed by graph instances in many fields, \eg in social networks). Hence, a large body of new research proposes n-ary (multiway) joins to achieve theoretically optimal complexity~\cite{DBLP:journals/jacm/NgoPRR18,DBLP:journals/pvldb/AmmarMSJ18}.
  
\item
  \label{item:untackled}
  \label{challenge:higher-order}
  \emph{Higher-order queries.}
  PG queries often employ \emph{higher-order} expressions \cite{DBLP:journals/tcs/BunemanNTW95}, \eg processing the vertices/edges on a path (also known as \emph{path unwinding}~\cite{DBLP:journals/csur/AnglesABHRV17}). 
  Incrementalization of higher-order languages is a new field of research~\cite{DBLP:conf/pldi/CaiGRO14}, and up to our best knowledge, currently there are no implementations using these techniques for query evaluation.
\end{enumerate}

In this paper, we address challenges
\ref{item:tackled-challenge-from}--\ref{item:tackled-challenge-to} in detail, present a first solution to handle
\ref{item:naive-challenge-from}--\ref{item:naive-challenge-to} with acceptable performance and propose techniques from the literature to tackle
\ref{item:future-challenge-from}--\ref{item:future-challenge-to}. Finding applicable techniques to handle \ref{item:untackled} is left for future work.

\fakeparagraph{Contributions}
In this paper, we discuss the challenges of IVM on PG queries and present an approach to tackle some of these challenges. In particular:

\begin{itemize}
\item We introduce extensions for relational algebra in order to handle graph-specific operators and use them to capture the semantics of (a subset of) the \opencypher language.
\item We define a mapping for \pg data using nested relations, and use nested relational algebra (NRA) to represent the queries. The data model can represent both the property graph and the resulting tables, while the NRA operators have sufficient expressive power to capture operations on the \pg. This allows the algebra to be \emph{composable and closed} even for operations such as transitive reachability.
\item We define a chain of transformations to translate the nested algebraic query plans to (incrementally maintainable) flat relational algebra (FRA) expressions.
\item We present the schema inferencing algorithm that eliminates the need to define the graph schema in advance.
\item We present \emph{\ingraph}, a research prototype capable of evaluating \opencypher graph queries incrementally.\footnote{\ingraph is available as an open-source tool at \url{http://github.com/ftsrg/ingraph}.}
\item We overview applicable IVM approaches from the literature in rule-based expert systems, integrity constraint checking, and materialized views.
\end{itemize}

\fakeparagraph{Paper structure}
We first present some theoretical background for property graphs (\autoref{sec:preliminaries}) and define the operators of graph relational algebra (\autoref{sec:gra}). We then discuss the compilation and query evaluation process (\autoref{sec:transformation}) and view maintenance (\autoref{sec:view-maintenance}).
Finally, we overview related techniques (\autoref{sec:related-work}) and outline future directions (\autoref{sec:summary}).

\section{The property graph data model}
\label{sec:preliminaries}

\setlength{\tabcolsep}{0.3em}
\begin{figure*}[t]
  \footnotesize
  \begin{subfigure}[b]{0.3\textwidth}
  \begin{align*}

  	\vertexlabels        & = \{\atom{Person}, \atom{Student}, \atom{Class}, \atom{Tag}\}                            \\
  	\edgelabels          & = \{\atom{KNOWS}, \atom{INTEREST}, \atom{SUBCLASS\_OF}, \atom{CLASS}\}                       \\
  	\vertexproperties    & = \{\atom{name}, \atom{speaks}, \atom{topic}, \atom{subject}\}, \quad                           
  	\edgeproperties      = \{\atom{since}\}                                                                       \\

    &\rule{\linewidth}{0.2mm} \\[0.15cm]
    
    V                    & = \{a, b, c, d, e, f\}, \quad                                                                   
  	E                    = \{1, 2, 3, 4, 5\}                                                                      \\
  	\verticestoedges     & :\ 1 \assign \tuple{a, b}, 
                              2 \assign \tuple{a, c}, \ldots \\
  	\vertexlabelfunction & :\ a \assign \{\atom{Person}, \atom{Student}\}, b \assign \{\atom{Person}\}, \ldots      \\
  	\edgelabelfunction   & :\ 1 \assign \atom{KNOWS}, 2 \assign \atom{INTEREST}, \ldots                             \\
  	\atom{name}          & :\ a \assign \atom{``Alice"}, b \assign \atom{``Bob"}, \ldots \atom{age} :\ a \assign 24,  \ldots \\
  	\atom{speaks}        & :\ a \assign \atom{\Lbag``en"\Rbag}, b \assign \atom{\Lbag``en", ``de"\Rbag}, c \assign \relnull, \ldots \\
  	\atom{since}         & :\ 1 \assign \atom{2014}, 2 \assign \relnull, \ldots \quad \atom{level}:\ 2 \assign 4, \ldots
  \end{align*}
  \caption{Example graph defined formally.}
  \label{fig:example-graph-formally}
  \end{subfigure}
  ~
  \begin{subfigure}[b]{0.3\textwidth}
  \scriptsize
  \centering
	\includegraphics[scale=0.38]{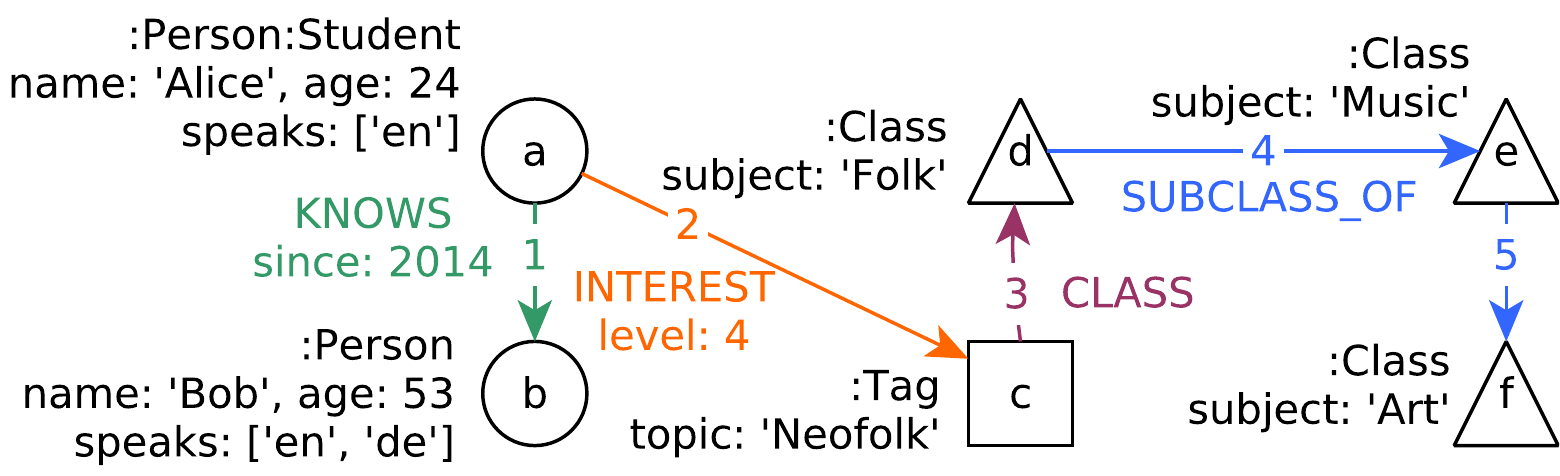}
	\caption{Example graph visualised.}
	\label{fig:example-graph-1}

 	\begin{tabular}{|c|c|c|l|P{0.82cm}|P{0.3cm}|}
    \hline
 		\multicolumn{6}{|c|}{\bf edge}                  \\ \hline
    \multirow{2}{*}{\bf id} &
    \multirow{2}{*}{\bf src} & 
    \multirow{2}{*}{\bf trg} &
    \multicolumn{1}{c|}{\multirow{2}{*}{\bf type}} & 
    \multicolumn{2}{c|}{\bf properties} \\ \cline{5-6}
    & & & & \bf key & \bf value \\ \hline
 		\edgerow{1}{a}{b}{KNOWS}{\propsi{since}{2014}} \\ \hline
 		\edgerow{2}{a}{c}{INTEREST}{\propsi{level}{4}} \\ \hline
 		\edgerow{3}{c}{d}{CLASS}{$-$} \\ \hline
 		\edgerow{4}{d}{e}{SUBCLASS\_OF}{$-$} \\ \hline
 		\edgerow{5}{e}{f}{SUBCLASS\_OF}{$-$} \\ \hline
 	\end{tabular}

  \caption{Nested relation of edges: $\gre$.}
  \label{fig:gre}

  \end{subfigure}
  ~
  \begin{subfigure}[b]{0.3\textwidth}
  \scriptsize
	\centering

 	\begin{tabular}{|c|c|P{0.82cm}|P{0.4cm}|}
    \hline
 		\multicolumn{4}{|c|}{\bf vertex} \\ \hline
    \multirow{2}{*}{\bf id} & \bf labels & \multicolumn{2}{c|}{\bf properties} \\ \cline{2-4}
    & \bf label & \bf key & \bf value\\ \hline
 		\vertexrow{a}{\labelsii{Student}{Person}}{\propsiii{name}{Alice}{age}{24}{speaks}{\itemsi{en}}} \\ \hline
 		\vertexrow{b}{\labelsi{Person}}{\propsiii{name}{Bob}{age}{53}{speaks}{\itemsii{en}{de}}} \\ \hline
 		\vertexrow{c}{\labelsi{Tag}}{\propsi{topic}{Neofolk}} \\ \hline
 		\vertexrow{d}{\labelsi{Class}}{\propsi{subject}{Folk}} \\ \hline
 		\vertexrow{e}{\labelsi{Class}}{\propsi{subject}{Music}} \\ \hline
 		\vertexrow{f}{\labelsi{Class}}{\propsi{subject}{Art}} \\ \hline
 	\end{tabular}

  \caption{Nested relation of vertices: $\grv$.}
  \label{fig:grv}
  \end{subfigure}
  
  \vspace{1.0ex}
  
  \begin{subfigure}[t]{0.23\textwidth}
  \scriptsize
 	\centering

 	\begin{tabular}{|c|c|P{0.82cm}|P{0.4cm}|}
  \hline
	\multicolumn{4}{|c|}{\bf $s$} \\ \hline
  \multirow{2}{*}{\bf id} & \bf labels & \multicolumn{2}{c|}{\bf properties} \\ \cline{2-4}
  & \bf label & \bf key & \bf value\\ \hline
 		\vertexrow{a}{\labelsii{Student}{Person}}{\propsiii{name}{Alice}{age}{24}{speaks}{\itemsi{en}}} \\ \hline
 	\end{tabular}

  \vspace{0.1ex}
  \caption{\emph{Get-vertices} result: $\left(\getvertices{s}{Student}\right)$.}
  
  \label{fig:getvertices-result}
  \end{subfigure}
  ~
  \begin{subfigure}[t]{0.68\textwidth}
  \scriptsize
  \centering
 	\begin{tabular}{|c|c|P{0.82cm}|P{0.4cm}|c|c|c|l|P{0.82cm}|P{0.3cm}|c|c|P{0.82cm}|P{0.4cm}|}
 		\hline
 		\multicolumn{4}{|c|}{\bf $s$} & \multicolumn{6}{c|}{\bf $i$} & \multicolumn{4}{c|}{\bf $t$} \\ \hline

    \multirow{2}{*}{\bf id} & \bf labels & \multicolumn{2}{c|}{\bf properties} &
    \multirow{2}{*}{\bf id} & \multirow{2}{*}{\bf src} & \multirow{2}{*}{\bf trg} & \multicolumn{1}{c|}{\multirow{2}{*}{\bf type}} & \multicolumn{2}{c|}{\bf properties} &
    \multirow{2}{*}{\bf id} & \bf labels & \multicolumn{2}{c|}{\bf properties} \\ \cline{2-4}\cline{9-10}\cline{13-14}

    & \bf label & \bf key & \bf value 
    & & & & & \bf key & \bf value 
    & & \bf label & \bf key & \bf value \\ \hline

  \vertexrow{a}{\labelsii{Student}{Person}}{\propsiii{name}{Alice}{age}{24}{speaks}{\itemsi{en}}} &
  \edgerow{2}{a}{c}{INTEREST}{\propsi{level}{4}} &
	\vertexrow{c}{\labelsi{Tag}}{\propsi{topic}{Neofolk}} \\ \hline
 	\end{tabular}
  \vspace{-1.2ex}
  \caption{A result relation produced by an application of the \emph{\getedgestext} operator: $\left[ \getedgesdirected{s}{Student}{t}{Tag}{i}{INTEREST} \right]$.}
  \label{fig:getedges-result}
  \end{subfigure}

	\caption{Social network example represented graphically, formally, and as nested relations.}
  \label{fig:example-graph}
\end{figure*}

\subsection{Data model}
\label{sec:pg-concepts}

The concept of the \emph{property graph} has only been studied by a few academic works, but it already has multiple flavours and definitions~\cite{DBLP:conf/edbt/HolschG16,DBLP:conf/adbis/MartonSV17,DBLP:conf/sigmod/AnglesABBFGLPPS18,DBLP:conf/sigmod/FrancisGGLLMPRS18,DBLP:conf/amw/Angles18}.
In this paper, we define it as follows.

\fakeparagraph{Structure}
A PG is $G = (V, E, \verticestoedges, \vertexlabels, \edgelabels, \vertexlabelfunction, \edgelabelfunction, \vertexproperties, \edgeproperties)$, where $V$ is a set of vertex identifiers, $E$ is a set of edge identifiers, and function $\verticestoedges: E \assign V \cartesianproductop V$ assigns the source and target vertices to edges. Vertices are labelled and edges are typed:
$\vertexlabels$ is a set of vertex labels, function $\vertexlabelfunction: V \assign 2^{\vertexlabels}$ assigns a \emph{set of labels} to each vertex;
$\edgelabels$ is a set of edge types, function $\edgelabelfunction: E \assign \edgelabels$ assigns a \emph{single type} to each edge.

\fakeparagraph{Properties}
Let $S$ be a set of scalar literals, and $\mathrm{FBAG}(X)$ denote the set of all finite bags of elements from $X$.
Let $D = S \unionop \mathrm{FBAG}(S)$ be the value domain for the \pg.
\footnote{The data model can be generalised further, \eg by allowing arbitrary nesting of collections. However, this data model already has higher expressive power than most graph data models (\eg semantic graphs) and satisfies the needs of most practical use cases. It is also powerful enough to represent the complex schema of the LDBC Social Network Benchmark~\cite{DBLP:conf/sigmod/ErlingALCGPPB15}.}

\begin{itemize}
\item $\vertexproperties$ is the set of vertex properties. A vertex property $p \in \vertexproperties$ is a partial function $p: V \assign D$, which assigns a property value $d \in D$ to a vertex $v \in V$, if $v$ has property $p$, otherwise $p(v)$ returns $\relnull$.

\item $\edgeproperties$ is the set of edge properties. An edge property $p \in \edgeproperties$ is a partial function $p: E \assign D$, which assigns a property value $d \in D$ to an edge $e \in E$, if $e$ has property $p$, otherwise $p(e)$ returns $\relnull$.
\end{itemize}

\fakeparagraph{Example graph}
An example graph inspired by the LDBC Social Network Benchmark~\cite{LDBC-SNB-BI-short} is shown formally in \autoref{fig:example-graph-formally} and graphically in \autoref{fig:example-graph-1}. The graph contains a $\mathsf{Tag}$, two $\atom{Persons}$, and three $\mathsf{TagClasses}$. Note that edges in the \pg data model are always \emph{directed}, hence the $\mathsf{KNOWS}$ relation is represented with a directed edge and the symmetric nature of the relation can be modelled in the queries.

\subsection{Nested relations}
\label{sec:nested-relations}
\opencypher queries take a \pglong as their inputs and return a \emph{graph relation}~\cite{DBLP:conf/edbt/HolschG16,DBLP:conf/adbis/MartonSV17} as their output.
To represent graphs and query results using the same algebraic constructs, we use \emph{nested relations}~\cite{DBLP:journals/is/Colby90}, which allow data items of a relation to contain additional relations with an arbitrary level of nesting. The domain for the internal relations is $D \union \{\relnull\}$. Relations on all levels of nesting follow bag semantics, \ie duplicate tuples are allowed. We define the schema of a relation as a \emph{list of (nested) attributes} and denote it with $\schema \left(r\right)$ for relation~$r$.

To represent the vertices and edges of the property graph, we define two nested relations, $\grv$ and $\gre$. Both relations
have a single attribute containing nested relations.
Their schema is given below and its mapping to the \pg concepts is in~\autoref{table:schema-vs-pg}.
$$\schema \left(\grv\right) = \schematuple{ \atom{vertex \big(id, labels(label), properties(key, value) \big) }} $$
$$\schema \left(\gre\right) = \schematuple{ \atom{edge   \big(id, src, trg, type,  properties(key, value) \big) }}$$
For $\grv.\atom{vertex}$, its $id$ corresponds to the elements in $V$. For a particular vertex, $\atom{labels}$ is the result of the $\vertexlabelfunction$ function, whereas $\atom{properties}$ is the result of $\vertexproperties$.
Similarly for $\gre.\atom{edge}$, $id$ corresponds to the elements in $E$. For a particular edge, the $\atom{type}$ corresponds to the result of $\edgelabelfunction$, $\atom{properties}$ is the result of $\edgeproperties$, and $\left(\mathit{src}, \mathit{trg}\right)$ is the result of $\verticestoedges$.

\begin{table}[t]
\renewcommand\arraystretch{1.1}
\begin{adjustbox}{center}
\scriptsize
\begin{tabular}{c|c|c|c}
	\toprule
    \multicolumn{2}{c|}{\bf vertex} & \multicolumn{2}{c}{\bf edge}  \\
    schema & \pg & schema &  \pg \\ \hline
    $$id$$ & $V$ & $$id$$ & $E$ \\
    $$labels(label)$$ & $\vertexlabelfunction$ & $$type$$ & $\edgelabelfunction$ \\
    $$properties$$ & $\vertexproperties$ & $$properties$$ & $\edgeproperties$ \\
    & & $\left(\mathit{src}, \mathit{trg}\right)$ attributes & $\verticestoedges$ \\

  \bottomrule
\end{tabular}
\end{adjustbox}
\caption{Mapping between the \pg data model and its representation as nested relations.}
\label{table:schema-vs-pg}
\end{table}

The nested relations representing the example graph are shown in Figure~\ref{fig:gre} and \ref{fig:grv}. These show that the set of vertex labels are stored as a nested relation $\atom{labels}$ with a single attribute $\atom{label}$, while edge types are simply stored as a single string value. The properties of vertices/edges are stored as a nested relation $\atom{properties}$ with two attributes, $\atom{key}$ and $\atom{value}$. This representation is well-suited to the flexible schema of \pg databases, as new labels, types, and property keys can be added without any changes to the schemas of the relations.

\section{Graph relational algebra}
\label{sec:gra}

Papers~\cite{DBLP:conf/edbt/HolschG16} and~\cite{DBLP:conf/adbis/MartonSV17} presented relational algebraic formalisations of the \opencypher language. A more rigorous formalisation was given in~\cite{DBLP:conf/sigmod/FrancisGGLLMPRS18}.
In this paper, we follow the approach of our previous work~\cite{DBLP:conf/adbis/MartonSV17} as it is better suited to established IVM techniques. This approach uses \emph{graph relational algebra} (GRA), which extends standard relational algebra operators with graph-specific navigation operators.

In this section, we formally define the operators of \gra and show example queries specified in natural language and as an \opencypher query, along with the equivalent \gra expression and the resulting output relation.

\subsection{Basic operators and nested property access}
\label{sec:property-access}

We first present the basic \emph{unary operators} of relational algebra, found in most relational algebra textbooks like~\cite{DBLP:books/daglib/0020812}.
The \emph{\selectiontext} operator $\selectionop$ filters the incoming relation according to some criteria.
Formally,
$ t = \selection{\theta} \left(r\right), $
where predicate $\theta$ is a propositional formula. Relation $t$ contains all tuples from $r$ for which $\theta$ holds. 
The \emph{\projectiontext} operator $\projectionop$ keeps a specific set of attributes in the relation: $ t = \projection{\var{x}_1, \ldots, \var{x}_n}{} \left(r\right).$ Note that the tuples are not deduplicated, thus $t$ will have the same number of tuples as $r$. The projection operator can also alias attributes, \eg $\projection{\var{x}_1 \alias \var{y}_1, 5 \alias \var{y}_2} \left(r\right)$ renames $\var{x}_1$ to $\var{y}_1$ and returns 5 as attribute $\var{y}_2$.
The \emph{\duplicateeliminationtext} operator $\duplicateeliminationop$ eliminates duplicate tuples in a bag, enforcing set semantics on its input.

\fakeparagraph{Shorthands}
For the sake of conciseness, we introduce two shorthands.
First, we allow using the dot notation ($.$) to traverse the nested schema to directly access nested attributes in the expressions (such as the selection predicate~$\theta$),
\eg the expression $\selection{\atom{vertex}.\atom{id} = a} \left(\grv\right)$ can access the $\atom{id}$ attribute of the attribute $\grv.\atom{vertex}$. This notation requires $\grv.\atom{vertex}$ to have an $\atom{id}$ and the expression holds iff $\atom{id}$ equals to $a$.

Second, \emph{properties} stored as key-value pairs in the nested $\atom{properties}$ relation can be accessed directly as if they were top-level attributes, 

\eg the expression $\selection{\atom{vertex}.\atom{age} = 25} \left(\grv\right)$ can access the $\atom{age}$ property of attribute $\grv.\atom{vertex}$. Unlike nested attribute access, this shorthand does not require $\grv.\atom{vertex}$ to have a property with key $\atom{age}$, it simply returns $\relnull$ in the absence of such a key.

\subsection{The get-vertices and get-edges operators}

For mapping a \pglong to relations, we use the nullary operators \emph{\getverticestext} and \emph{\getedgestext}. We define these operators using the nested relations $\grv$ and $\gre$ introduced in \autoref{sec:nested-relations}. These operators are rather involved, hence we introduce some notational conventions used for the definitions:

\begin{itemize}
\item A vertex variable $v$ is \emph{free} w.r.t.\ an operator's input relation $r$ if $v \not\in \schema \left( r \right)$ and bound if $v \in \schema \left( r \right)$.
\item $\vertexfree$ represents a free vertex, $\vertexbound$ represents a bound vertex, and $\vertexmaybe$ represents any vertex.
\item Arrow symbols $\rightarrow$, $\leftarrow$, and $\leftrightarrow$ represent an outgoing, incoming, and undirected edge, respectively.
\item
For vertices, we use three predefined sets of labels:\\
$\atom{\labelsL} \equiv \atom{\labelsLlong}$;
$\atom{\labelsA} \equiv \atom{\labelsAlong}$; and
$\atom{\labelsB} \equiv \atom{\labelsBlong}$.
\item For edges, we use a set of types $\atom{\types} \equiv \atom{\typeslong}$.
\end{itemize}

\fakeparagraph{Get-vertices}
The \emph{\getverticestext} operator~\cite{DBLP:conf/edbt/HolschG16} $\left( \getvertices{v}{\labelsL} \right)$ returns a nested relation of a single attribute $v$ that contains vertices which have \emph{all} labels of $\atom{\labelsL}$. Formally, it is defined as:
$$
\left(
\getvertices{v}{\labelsL}
\right)
\equiv
\projection{\grv.\atom{vertex}/\var{v}}
\left(
\selection{\atom{\labelsL} \subseteq \atom{\grv.vertex.labels}} \left(\grv\right)
\right)
$$

The schema of the resulting relation is $\schema \left( \getvertices{v}{\labelsL} \right) \equiv \schematuple{v}$, as the example in \autoref{fig:getvertices-result} shows. The usage of the operator is illustrated with the following example:

  \examplespacing
  \noindent\begin{minipage}{\linewidth}
  \rule{\linewidth}{0.8pt}
  {\bf Example.} {\it Get the name of all Persons aged over 25.}
  
  \vspace{-1.5ex}\rule{\linewidth}{0.4pt}
  
  \noindent\lstinputlisting{queries/getvertices.cypher}
  
  \vspace{-2ex}\rule{\linewidth}{0.4pt}\examplespacing

  \centering$
\projection{\var{p}.\atom{name}}
\selection{\var{p}.\atom{age} > 25}
\left(
\getvertices{p}{Person}
\right)
$
  
  \rule{\linewidth}{0.4pt}\examplespacing
  
  \centering
  \scriptsize
\begin{tabular}{|l|}
\hline
\multicolumn{1}{|c|}{\bf $\var{p}.\atom{name}$} \\ \hline
Bob \\ \hline
\end{tabular}

  \examplespacing\rule{\linewidth}{0.8pt}
  \end{minipage}
  
  \examplespacing\examplespacing\examplespacing
{The get-edges operator}
Next, we introduce the \emph{\getedgestext} operator $\left[ \getedgesopdirected \right]$, which returns edges along with their source and target vertices. Using theta joins on \emph{\getverticestext} operators and relation~$\gre$, the \getedgestext operator can be defined as:
\begin{align*}
&\left[ \getedgesdirected{v}{\labelsA}{w}{\labelsB}{e}{\types} \right]
\equiv
\projection{v, e, w} \Big(
\\
&
\quad
\left(\getvertices{v}{\labelsA}\right)
\underset{\var{v}.\atom{id} = \atom{\gre.edge.src}}{\join}
\big(\selection{\atom{\gre.edge.type} \,\in\, \atom{\types}} \left(\gre\right)\big)
\underset{\atom{\gre.edge.trg} = \var{w}.\atom{id}}{\join}
\left(\getvertices{w}{\labelsB}\right)
\Big)
\end{align*}

The schema of the result is $\schema \left[ \getedgesdirected{v}{\labelsA}{w}{\labelsB}{e}{\types} \right] \equiv  \schematuple{v, e, w}$, as the example in \autoref{fig:getedges-result} shows.

\fakeparagraph{Edge directions}
Additionally to the directed \getedgestext operator, we define the \emph{undirected \getedgestext} operator~$\getedgesopundirected$, which enumerates edges of both directions. Formally:
$$
\left[ \getedgesundirected{v}{\labelsA}{w}{\labelsB}{e}{\types} \right]
\equiv
\left[ \getedgesdirected{v}{\labelsA}{w}{\labelsB}{e}{\types} \right]
\,\Union\,
\projection{v, e, w}
\left[ \getedgesdirected{w}{\labelsB}{v}{\labelsA}{e}{\types} \right]
$$

\fakeparagraph{Notation}
To aid readability, we always surround the $\left(\getvertices{v}{L}\right)$ and $\Big[\getedgesdirected{v}{\labelsA}{w}{\labelsB}{e}{T}\Big]$ operators with parentheses and brackets, resp.

\subsection{The expand operators}
\label{sec:expand}

To capture navigations, we define the unary \emph{\expandouttext} operator~$\expandoutop$. The expression $\expandout{v}{w}{\labelsL}{e}{\types}{1}{1} (r)$ takes tuples from relation $r$ and returns a tuple for each possible navigation from a bound vertex $v$ to vertex $w$ through an edge $e$, while enforcing the label and type constraints ($w$ is labelled with all labels of $\atom{\labelsL}$ and $e$ is typed with one type of $\atom{\types}$ or has any type if $T$ is empty). It can be defined using the \emph{\getedgestext} operator:
$$
\expandout{v}{w}{\labelsL}{e}{\types}{1}{1} \left(r\right) =
r \join
\left[ \getedgesdirected{v}{}{w}{\labelsL}{e}{\types} \right]
$$

The schema of the resulting relation is $\schema \left( \expandout{v}{w}{\labelsL}{e}{\types}{1}{1} \right) \equiv \schema \left( r \right) \union \schematuple{e, w}$. 
The operator is demonstrated as follows:

  \examplespacing
  \noindent\begin{minipage}{\linewidth}
  \rule{\linewidth}{0.8pt}
  {\bf Example.} {\it Get Persons and their interests.}
  
  \vspace{-1.5ex}\rule{\linewidth}{0.4pt}
  
  \noindent\lstinputlisting{queries/expandout.cypher}
  
  \vspace{-2ex}\rule{\linewidth}{0.4pt}\examplespacing

  \centering$
\projection{\var{p}.\atom{name}, \var{i}.\atom{level}, \var{t}.\atom{topic}}
\left(\expandout{p}{t}{Tag}{i}{INTEREST}{1}{1}
\left(\getvertices{p}{Person}\right)\right)
$

  \rule{\linewidth}{0.4pt}\examplespacing
  
  \centering
  \scriptsize
\begin{tabular}{|l|l|l|}
\hline
\multicolumn{1}{|c|}{\bf $\var{p}.\atom{name}$} &
\multicolumn{1}{c|}{\bf $\var{i}.\atom{level}$} &
\multicolumn{1}{c|}{\bf $\var{t}.\atom{topic}$}
\\\hline
Alice & 4 & Neofolk \\\hline
\end{tabular}

  \examplespacing\rule{\linewidth}{0.8pt}
  \end{minipage}
  
  \examplespacing\examplespacing\examplespacing
{Edge directions}
We define two additional \emph{\expandtext} operators: 
the \emph{\expandintext} operator~$\expandinop$ accepts incoming edges,
while
the \emph{\expandbothtext} operator~$\expandbothop$ accepts edges from both directions. Formally, they can be defined as follows:
$$
\expandin{v}{w}{\labelsL}{e}{\types}{1}{1} \left(r\right)
\equiv 
r
\join
\left[
\getedgesdirected{w}{\labelsL}{v}{}{e}{\types}
\right],
\quad
\expandboth{v}{w}{\labelsL}{e}{\types}{1}{1} \left(r\right)
\equiv 
r
\join
\left[
\getedgesundirected{v}{}{w}{\labelsL}{e}{\types}
\right]
$$

\fakeparagraph{Transitive navigation}
To allow multi-hop navigation along the edges, we define a transitive variant of the expand operator $\expandout{v}{w}{B}{E}{T}{low}{up}$, which navigates from $v$ to $w$ through edges~$E$ of any type in $\atom{\types}$ (if $\atom{\types}$ is not empty), using a number of hops between a lower bound ($\atom{low}$) and an upper bound ($\atom{up}$).

We restate here that the nested relations in this paper follow bag semantics (\autoref{sec:nested-relations}), \ie they do not store any ordering between their tuples.
Therefore, storing the edges of a paths as a single attribute would cause us to lose the information on ordering.
Therefore, we define attribute~$E$ as a nested attribute which stores the edge attribute ``$\atom{edge}$'' along with an indexing attribute ``$\atom{index}$'' that denotes the position of the edge in the path.
Using this attribute, the schema is:
$$\schema \left( \expandout{v}{w}{\labelsL}{E}{\types}{low}{up} \left(r\right) \right) \equiv \schema \left( r \right) \union \schematuple{E\left(\atom{index}, \atom{edge}\right), w}$$

This is demonstrated with the following example:

  \examplespacing
  \noindent\begin{minipage}{\linewidth}
  \rule{\linewidth}{0.8pt}
  {\bf Example.} {\it Get the subclasses of Class 'Art'}
  
  \vspace{-1.5ex}\rule{\linewidth}{0.4pt}
  
  \noindent\lstinputlisting{queries/multihop.cypher}
  
  \vspace{-2ex}\rule{\linewidth}{0.4pt}\examplespacing

  \centering$
\projection{\var{c}.\atom{name}, \var{sos}}
\left(
\expandin{a}{c}{Class}{sos}{SUBCLASS\_OF}{1}{\infty}
\Big(
\selection{\var{a}.\atom{topic} = \atom{'Art'}}
\left(\getvertices{a}{Class}\right)
\Big)
\right)
$

  \rule{\linewidth}{0.4pt}\examplespacing
  
  \centering
  \scriptsize
\begin{tabular}{|c|c|c|c|c|l|l|l|}
\hline
\multirow{4}{*}{\bf $\var{c}.\atom{name}$} &
\multicolumn{7}{c|}{\bf $\var{sos}$}
\\ \cline{2-8}
&\multirow{3}{*}{\bf index} &
\multicolumn{6}{c|}{\bf edge}                  \\ \cline{3-8}
& &
\multirow{2}{*}{\bf id} &
\multirow{2}{*}{\bf src} & 
\multirow{2}{*}{\bf trg} &
\multicolumn{1}{c|}{\multirow{2}{*}{\bf type}}
&\multicolumn{2}{c|}{\bf properties} \\ \cline{7-8}
& & & & & & \bf key & \bf value \\ \hline
\multirow{2}{*}{Folk} &	1 & \edgerow{4}{d}{e}{SUBCLASS\_OF}{$-$} \\ \cline{2-8}
                      &	2 & \edgerow{5}{e}{f}{SUBCLASS\_OF}{$-$} \\ \hline
Music &	1 & \edgerow{5}{e}{f}{SUBCLASS\_OF}{$-$} \\ \hline
\end{tabular}


  \examplespacing\rule{\linewidth}{0.8pt}
  \end{minipage}
  
  \examplespacing\examplespacing\examplespacing
{Combining pattern matches}

A single graph pattern is defined starting from \emph{\getverticestext} and \emph{\expandtext}  operators. Multiple graph patterns can be combined together based on their common attributes using the natural join operator~$\joinop$. Additionally, most \pg query languages allow users to define optional pattern parts. This can be captured with the \emph{\leftouterjointext} operator $\leftouterjoin$, which pads tuples from the left relation that did not match any from the right relation with $\relnull$ values and adds them to the result of the \jointext~\cite{DBLP:books/daglib/0015084}.
This is illustrated by the following example:

  \examplespacing
  \noindent\begin{minipage}{\linewidth}
  \rule{\linewidth}{0.8pt}
  {\bf Example.} {\it Get Persons and their interests if they have any.}
  
  \vspace{-1.5ex}\rule{\linewidth}{0.4pt}
  
  \noindent\lstinputlisting{queries/loj.cypher}
  
  \vspace{-2ex}\rule{\linewidth}{0.4pt}\examplespacing

  \centering$
\projection{\var{p}.\atom{name}, \var{t}}
\left(
\left(\getvertices{p}{Person}\right)
\leftouterjoin
\left[\getedgesdirected{p}{}{t}{Tag}{i}{INTEREST}\right]
\right)
$
  
  \rule{\linewidth}{0.4pt}\examplespacing
  
  \centering
  \scriptsize
\begin{tabular}{|l|c|c|P{1.09cm}|P{1.2cm}|}
	\hline
	\multicolumn{1}{|c|}{\multirow{3}{*}{\bf $\var{p}.\atom{name}$}} &                                          \multicolumn{4}{c|}{\bf $t$}                                           \\ \cline{2-5}
	                                                  &                  \multirow{2}{*}{\bf id}                  & \bf labels & \multicolumn{2}{c|}{\bf properties} \\ \cline{3-5}
	                                                  &                                                           & \bf label  & \bf key & \bf value                 \\ \hline
	Alice                                             & \vertexrow{c}{\labelsi{Tag}}{\propsi{topic}{Neofolk}} \\ \hline
  Bob & \multicolumn{4}{c|}{$\relnull$} \\ \hline
\end{tabular}

  \examplespacing\rule{\linewidth}{0.8pt}
  \end{minipage}
  
  \examplespacing\examplespacing\examplespacing
ome queries pose structural conditions on the graph patterns (\eg only return Persons who have at least one interest). Positive structural conditions can be captured with the \emph{\semijointext} operator~$\semijoin$, which is defined as $r \semijoin s \equiv \projection{\schema \left( r \right)} \left(r \join s\right)$. Negative structural conditions can be captured by using the \emph{\antijointext} operator~$\antijoin$ (also known as the \emph{anti-semijoin}), which is defined as $r \antijoin s \equiv r - \left(r \semijoin s\right) $. For the sake of brevity, we refrain from providing examples for these operators.

\subsection{Collections and aggregation}

\fakeparagraph{Unwinding}
It is often required to handle elements in nested collections separately. To allow this, we introduce the \emph{\unwindtext} operator~$\unwindop$, a specialized version of the unnest operator~$\mu$ of nested relational algebra~\cite{botoeva_et_al:LIPIcs:2018:8607}.
In particular, $\unwind{xs \unwindarrow x}\left(r\right)$ takes the bag in attribute $\mathit{xs}$ and creates a new tuple for each element of the bag by appending that element as an attribute $x$ to $r_i \in r$.

\fakeparagraph{Ordering}
In common extensions to relational algebra~\cite{DBLP:books/daglib/0020812}, the \emph{\sorttext} operator $\sortop$ is used to sort a relation, returning a relation that follows \emph{list semantics}. The ordering is defined according to selected attributes and with a certain direction for each attribute (ascending~$\asc$ or descending~$\desc$), \eg $\sortop_{\asc \var{x}_1, \desc \var{x}_2} (r)$.
Additionally, the \emph{\toptext} operator $\topp{l}{s}$~\cite{DBLP:conf/sigmod/LiCIS05} takes a list relation as its input, skips the first $s$ tuples and returns the next $l$ tuples. The default values are 0 for $s$ and $\infty$ for $l$.

\newcommand{\vars}{\var{v_1}, \ldots, \var{v_n}}

As the operators in our nested bag algebra do not define ordering, a standalone \emph{\sorttext} or \emph{\toptext} operator would have no clear semantics. Hence, we only allow these operator combined together as a single \emph{\sortandtoptext} operator.
$$
\topp{\atom{l}}{\atom{s}} \left(\sort{\vars} \left(r\right) \right)
\Rightarrow
\sortandtop{\vars}{\atom{l}}{\atom{s}} \left(r\right)
$$

\fakeparagraph{Grouping and aggregation}
The \emph{\groupingtext} operator $\groupingop$ groups tuples according to their value in one or more attributes and aggregates the remaining attributes.
As determining the attributes of the \emph{grouping criteria} is non-trivial, the \emph{\groupingtext} operator explicitly states these attributes.
We use the notation $\grouping{e_1 \alias a_1, \ldots, e_n \alias a_n}{c_1, \ldots, c_n}$, where $c_1, \ldots, c_n$ form the \emph{grouping criteria}, \ie the list of expressions whose values partition the incoming tuples into groups. For every group this aggregation operator emits a single tuple of expressions $\tuple{e_1, \ldots, e_n}$ with aliases $\tuple{a_1, \ldots, a_n}$, respectively.
We demonstrate the \unwindtext, \groupingtext, \sorttext, and \toptext operators using a single example:

  \examplespacing
  \noindent\begin{minipage}{\linewidth}
  \rule{\linewidth}{0.8pt}
  {\bf Example.} {\it Number of speakers of the top 1 spoken language.}
  
  \vspace{-1.5ex}\rule{\linewidth}{0.4pt}
  
  \noindent\lstinputlisting{queries/grouping.cypher}
  
  \vspace{-2ex}\rule{\linewidth}{0.4pt}\examplespacing

  \centering$
\sortandtop{\desc \var{sks}}{1}{}
\left(
\grouping{\var{lang}, \texttt{count} \left( \var{p} \right)	\assign \var{sks}}{\var{lang}}
\left(
\unwind{\var{p}.\atom{speaks} \unwindarrow \var{lang}}
\left(
\getvertices{p}{Person}
\right)
\right)
\right)
$

  \rule{\linewidth}{0.4pt}\examplespacing
  
  \centering
  \scriptsize
\begin{tabular}{|l|l|}
	\hline
  \multicolumn{1}{|c|}{\bf $\var{lang}$} & \multicolumn{1}{c|}{\bf $\var{sks}$} \\ \hline
  en & 2 \\ \hline
\end{tabular}

  \examplespacing\rule{\linewidth}{0.8pt}
  \end{minipage}
  
  \examplespacing\examplespacing\examplespacing
{table}[t]
\setlength\extrarowheight{3.5pt}
\setlength\tabcolsep{3.6pt}
\begin{adjustbox}{center}
\scriptsize
\begin{tabular}{@{}ll@{}}
	\toprule
	\bf \small Language construct                                                                                                          & \bf \small GRA expression                                                                               \\ \midrule

	\lstinline+(<<v>>)+                                                                                                             & $\getvertices{v}{}$                                                                                          \\
	\lstinline+(<<v>>:<<l1>>:...:<<lk>>)+                                                                                           & $\getvertices{v}{\labelsLlong}$                                                                  \\
	\lstinline+<p>p</p>-[<<e>>:<<t1>>|...|<<to>>]->(<<w>>)+                                                                         & $\expandout{v}{w}{}{e}{\typeslong}{1}{1} (p)$                                                    \\
	\lstinline+<p>p</p><-[<<e>>:<<t1>>|...|<<to>>]-(<<w>>)+                                                                         & $\expandin{v}{w}{}{e}{\typeslong}{1}{1} (p)$                                                     \\
	\lstinline+<p>p</p><-[<<e>>:<<t1>>|...|<<to>>]->(<<w>>)+                                                                        & $\expandboth{v}{w}{}{e}{\typeslong}{1}{1} (p)$                                                   \\
	\lstinline+<p>p</p>-[<<E>>:<<t1>>|...|<<to>>*low..up]->(<<w>>)+                                                                          & $\expandout{v}{w}{}{E}{\typeslong}{low}{up} (p)$                                                                      \\ \midrule

	\lstinline+MATCH <p>p1</p>, <p>p2</p>, ...+                                                                                     & $ p1 \join p_2 \join \ldots $                              \\
	\lstinline+OPTIONAL MATCH <p>p</p>+                                                                                             & $ \dual \ \leftouterjoin\ p$                                    \\
	\lstinline+[[r]] OPTIONAL MATCH <p>p</p>+                                                                                       & $r \ \leftouterjoin\ p$          \\

	\lstinline+[[r]] WHERE <<condition>>+                                                                                                & $\selection{\atom{condition}}{\left( r \right)}$                                                                  \\
	\lstinline+[[r]] WHERE (<<v>>:<<l1>>:...:<<lk>>)+                                                                               & $\selection{\{\labelsLlong\} \subseteq \vertexlabelfunction(v)} (r) $  \\
	\lstinline+[[r]] WHERE <p>p</p>+                                                                                                & $r \semijoin p$                                                                                            \\
	\lstinline+[[r]] WHERE NOT <p>p</p>+                                                                                            & $ r \antijoin p $                                                                                            \\
	\lstinline+[[r]] RETURN <<x1>> AS <<y1>>, ...+                                                                                  & $\projection{x_1 / y_1, \ldots} \left( r \right)$                                        \\
	\lstinline+[[r]] RETURN DISTINCT <<x1>> AS <<y1>>, ...+                                                                         & $\duplicateelimination\left(\projection{x_1 / y_1, \ldots} \left( r \right)\right)$      \\
	\lstinline+[[r]] RETURN <<x1>>, <<x2>>, aggr(<<x3>>)+                                                                               & $\grouping{x_1, x_2, \atom{aggr}(x_3)}{x_1, x_2}{\left( r \right)}$                                  \\

	\lstinline+[[r]] UNWIND <<xs>> AS <<x>>+                                                                                        & $\unwind{\var{xs} \unwindarrow x}{\left( r \right)}$                                                                    \\
	\lstinline+[[r]] ORDER BY <<x1>> ASC, <<x2>> DESC, ...+                                                                         & \multirow{2}{*}{$\sortandtop{\asc x_1, \desc x_2, \ldots}{\atom{l}}{\atom{s}} {\left( r \right)}$}                                           \\[-1ex]
	\lstinline+    SKIP s LIMIT l+                                                                                        & \\

  \bottomrule
\end{tabular}
\end{adjustbox}
\caption{Mapping from \opencypher constructs to GRA. Variables, labels, and types are typeset as \mbox{\lstinline|<<v>>|}. The notation \mbox{\lstinline|<p>p</p>|} represents a pattern resulting in a relation $p$. To allow navigation from this relation, we presume that relation $p$ has an attribute $v$ that represents a vertex. \mbox{\lstinline|[[r]]|} stands for a relation $r$ that is a results of the previous query parts. To avoid confusion with the ``\lstinline+..+'' language construct (used for ranges), we use ``\lstinline+...+'' to denote omitted query parts.}
\label{table:mapping}
\end{table}

In this section, we defined the operators of \gra and gave an informal specification for compiling from \opencypher queries.
\autoref{table:mapping} shows a compact mapping of \opencypher queries to \gra expressions.
Note that the \emph{\getedgestext} operator is not needed to capture the mapping---instead, only the \emph{\getverticestext} nullary operators are used and edges are inserted by the \emph{\expandtext} and \emph{\transitiveexpandtext} operators.
For a more detailed mapping, we refer the reader to~\cite{DBLP:conf/adbis/MartonSV17}.

\section{Transforming Graph RA to Flat RA}
\label{sec:transformation}

\lstset{numbers=left,frame=single}

\begin{figure*}[h]
	\centering
  \begin{minipage}{\linewidth}
  \begin{minipage}[b]{0.25\textwidth}
	\begin{subfigure}[]{\textwidth}
  \centering
  \includegraphics[scale=\yedscale]{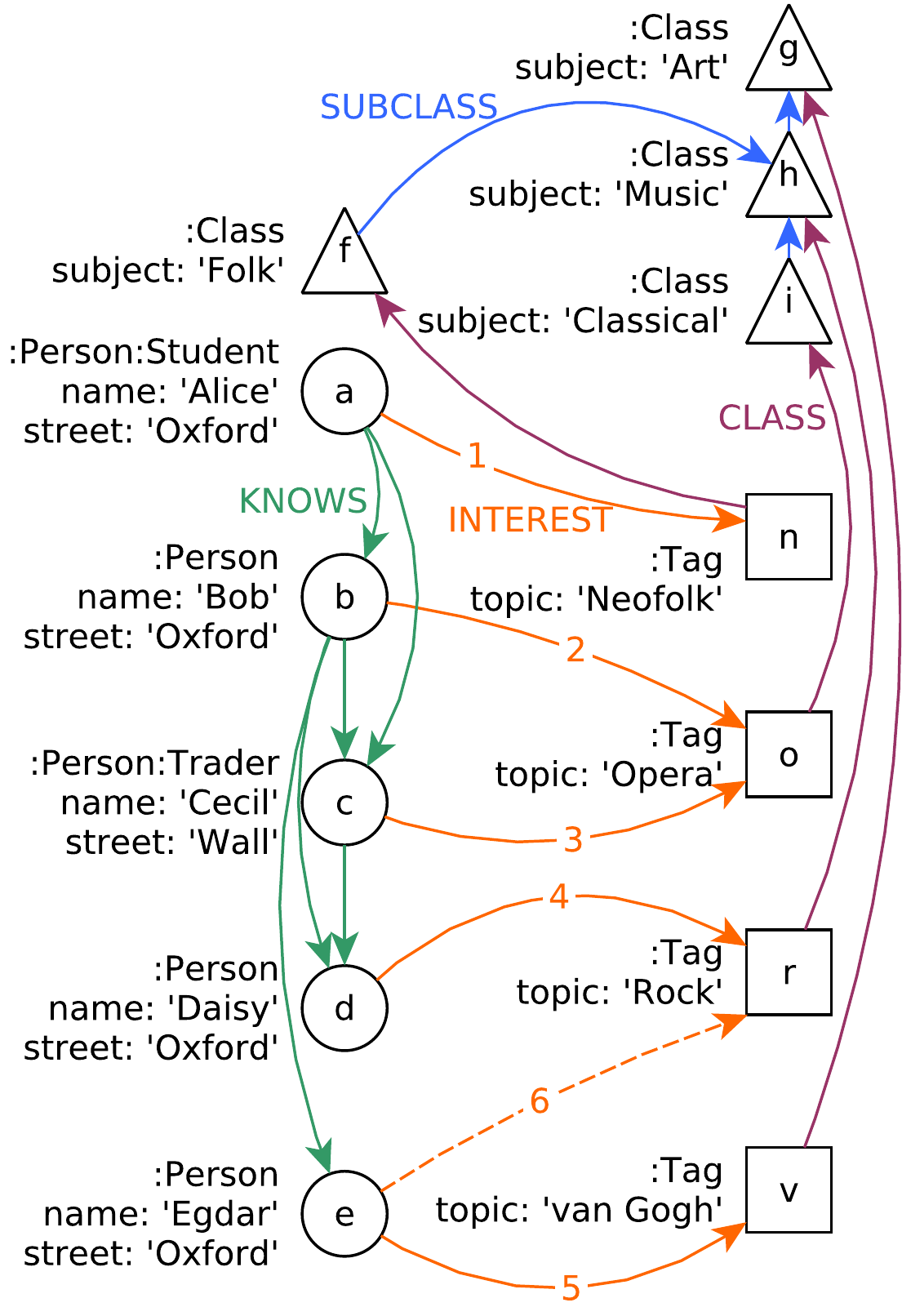}
  \caption{Example graph.}
  \label{fig:example-graph-2}
  \vspace{0.5ex}
  \end{subfigure}
	\begin{subfigure}[b]{\textwidth}
  \centering
  \footnotesize
\begin{tikzpicture}[->,>=stealth',node distance=0.2cm]
\tikzset{initial text={}}
\tikzstyle{every state}=[shape=rectangle,minimum width=3.4cm,text width=3.3cm,minimum height=0.33cm,align=center]

\node[state] (QE)               {Query specification \hfill \autoref*{fig:query-specification}};
\node[state] (QP) [below=of QE] {\gra query plan \hfill \autoref*{fig:query-plan-search}};
\node[state] (JP) [below=of QP] {\nra query plan \hfill \textcolor{nestedschemacolor}{\autoref*{fig:query-plan-rete}}};
\node[state] (FP) [below=of JP] {\fra query plan \hfill \textcolor{flatschemacolor}{\autoref*{fig:query-plan-rete}}};
\node[state] (QD) [below=of FP] {Query view \hfill \autoref{sec:view-maintenance}};

\path (QE) edge [] (QP)
      (QP) edge [] (JP)
      (JP) edge [] (FP)
      (FP) edge [] (QD)
;
\end{tikzpicture}
  \caption{The workflow of \ingraph.}
  \label{fig:workflow}
  \end{subfigure}
  \end{minipage}~
  \begin{minipage}[b]{0.3\textwidth}
	\begin{subfigure}{1.25\textwidth}
\begin{lstlisting}[label=lst:running-example,basicstyle=\footnotesize\ttfamily]
MATCH (p:Person)-[pi:INTEREST]->(pt:Tag)-[tc:CLASS]->
      (fc:Class)-[sos:SUBCLASS_OF*0..]->(c:Class)
WHERE c.subject = 'Music'
OPTIONAL MATCH (p)-[k:KNOWS]-(f:Person) 
WHERE p.street = f.street
WITH p, count(DISTINCT f) AS nf WHERE nf < 3
RETURN p.name
\end{lstlisting}
    \vspace{-0.5ex}
    \caption{Query specification in \opencypher.}
    \label{fig:query-specification}
    \vspace{1ex}
	\end{subfigure}
  \begin{subfigure}[b]{\columnwidth}
		\centering
    \hspace{-3ex}
    \tiny
    \togglefalse{textualoperators}
\begin{forest} for tree={align=center}
[
				{$\projection{\var{p}.\atom{name}}
				$
				}[
				{$\selection{
					\var{nf} < \literal{3}
				}
				$
				}[
				{$\grouping{\var{p}, \texttt{count\ distinct}
				\left( \var{f} \right)
				\assign \var{nf}}{\var{p}}$
				}[
				{$\leftouterjoin \{\var{p}\}$
				}[
				{$\selection{
					\var{c}.\atom{subject} = \literal{"Music"}
				}
				$
				}[
				{$\expandout{fc}{c}{Class}{sos}{SUBCLASS\_OF}{0}{\infty}$
				}[
				{$\expandout{pt}{fc}{Class}{tc}{CLASS}{1}{1}$
				}[
				{$\expandout{p}{pt}{Tag}{pi}{INTEREST}{1}{1}$
				}[
				{$\getvertices{p}{Person}$
				}]
]
]
]
]
[
				{$\selection{
					\var{p}.\atom{street} = \var{f}.\atom{street}
				}
				$
				}[
				{$\expandboth{p}{f}{Person}{k}{KNOWS}{1}{1}$
				}[
				{$\getvertices{p}{Person}$
				}]
]
]
]
]
]
]
;
\end{forest}
		\caption{Query plan in graph relational algebra.}
    \label{fig:query-plan-search}
	\end{subfigure}%
  \end{minipage}~
	\begin{subfigure}[b]{0.4\textwidth}

    \hspace{-6ex}
    \tiny
    \togglefalse{textualoperators}
\begin{forest} for tree={align=left}
[
				{$\projection{\var{p}.{\atom{name}}}
				$
				\\
				$\color{nestedschemacolor} \langle \var{p}.\atom{name} \rangle $
				\\
					$\color{flatschemacolor} \langle \uline{\var{p}.\atom{name}}
					 \rangle$
				}[
				{$\selection{
					\var{nf} < \literal{3}
				}
				$
				\\
				$\color{nestedschemacolor} \langle \var{p} , \var{nf} \rangle $
				\\
					$\color{flatschemacolor} \langle \var{\var{p} }
					, \var{\var{nf}}
					, \uline{\var{p}.\atom{name}}
					 \rangle$
				}[
				{$\grouping{\var{p}, \texttt{count\ distinct}
				\left( \var{f} \right)
				\assign \var{nf}, \var{p}.\atom{name}}{\var{p}, \var{p}.\atom{name}}$
				\\
				$\color{nestedschemacolor} \langle \var{p} , \var{nf} \rangle $
				\\
					$\color{flatschemacolor} \langle \var{\var{p} }
					, \var{\var{nf}}
					, \uline{\var{p}.\atom{name}}
					 \rangle$
				}[
				{$\leftouterjoin \{\var{p}\}$
				\\
				$\color{nestedschemacolor} \langle \var{p} , \var{pi} , \var{pt} , \var{tc} , \var{fc} , \var{sos}, \var{c} , \var{f} , \var{k}  \rangle $
				\\
					$\color{flatschemacolor} \langle \var{\var{p} }
					, \var{\var{pi} }
					, \var{\var{pt} }
					, \uline{\var{p}.\atom{name}}
					, \var{\var{tc} }
					, \var{\var{fc} }
					, \var{\var{sos}}
					, \var{\var{c} }
					, \uline{\var{c}.\atom{subject}}
					, \var{\var{f} }
					, \var{\var{k} }
					 \rangle$
				}[
				{$\selection{
					\var{c}.\atom{subject} = \literal{"Music"}
				}
				$
				\\
				$\color{nestedschemacolor} \langle \var{p} , \var{pi} , \var{pt} , \var{tc} , \var{fc} , \var{sos}, \var{c}  \rangle $
				\\
					$\color{flatschemacolor} \langle \var{\var{p} }
					, \var{\var{pi} }
					, \var{\var{pt} }
					, \uline{\var{p}.\atom{name}}
					, \var{\var{tc} }
					, \var{\var{fc} }
					, \var{\var{sos}}
					, \var{\var{c} }
					, \uline{\var{c}.\atom{subject}}
					 \rangle$
				}[
				{$\transitivejoin{0}{\infty} \{\var{fc}\}$
				\\
				$\color{nestedschemacolor} \langle \var{p} , \var{pi} , \var{pt} , \var{tc} , \var{fc} , \var{sos}, \var{c}  \rangle $
				\\
					$\color{flatschemacolor} \langle \var{\var{p} }
					, \var{\var{pi} }
					, \var{\var{pt} }
					, \uline{\var{p}.\atom{name}}
					, \var{\var{tc} }
					, \var{\var{fc} }
					, \var{\var{sos}}
					, \var{\var{c} }
					, \uline{\var{c}.\atom{subject}}
					 \rangle$
				}[
				{$\join \{\var{pt}\}$
				\\
				$\color{nestedschemacolor} \langle \var{p} , \var{pi} , \var{pt} , \var{tc} , \var{fc}  \rangle $
				\\
					$\color{flatschemacolor} \langle \var{\var{p} }
					, \var{\var{pi} }
					, \var{\var{pt} }
					, \uline{\var{p}.\atom{name}}
					, \var{\var{tc} }
					, \var{\var{fc} }
					 \rangle$
				}[
				{$\getedgesdirected{p}{Person}{pt}{Tag}{pi}{INTEREST}$
				\\
				$\color{nestedschemacolor} \langle \var{p} , \var{pi} , \var{pt}  \rangle $
				\\
					$\color{flatschemacolor} \langle \var{\var{p} }
					, \var{\var{pi} }
					, \var{\var{pt} }
					, \uline{\var{p}.\atom{name}}
					 \rangle$
				},for tree={}]
[
				{$\getedgesdirected{pt}{Tag}{fc}{Class}{tc}{CLASS}$
				\\
				$\color{nestedschemacolor} \langle \var{pt} , \var{tc} , \var{fc}  \rangle $
				\\
					$\color{flatschemacolor} \langle \var{\var{pt} }
					, \var{\var{tc} }
					, \var{\var{fc} }
					 \rangle$
				},for tree={}]
]
[
				{$\getedgesdirected{fc}{}{c}{Class}{sos}{SUBCLASS\_OF}$
				\\
				$\color{nestedschemacolor} \langle \var{fc} , \var{sos} , \var{c}  \rangle $
				\\
					$\color{flatschemacolor} \langle \var{\var{fc} }
					, \var{\var{sos} }
					, \var{\var{c} }
					, \uline{\var{c}.\atom{subject}}
					 \rangle$
				},for tree={}]
]
]{}
[
				{$\getedgesundirected{f}{Person}{p}{Person}{k}{KNOWS}$
				\\
				$\color{nestedschemacolor} \langle \var{f} , \var{k} , \var{p}  \rangle $
				\\
					$\color{flatschemacolor} \langle \var{\var{f} }
					, \var{\var{k} }
					, \var{\var{p} }
					 \rangle$
				},for tree={}]
]
]
]
]
;
\end{forest}
    \renewcommand{\thesubfigure}{f}
    \llap{\makebox[1.9cm][r]{\raisebox{1cm}{{
	\footnotesize
  \begin{minipage}{1.7cm}
  \centering
 	\begin{tabular}{|l|}
 		\hline
 		\bf p.name \\ \hline\hline
 		Alice \\ \hline
 		Cecil \\ \hline
 		Daisy \\ \hline \hdashline
    \multicolumn{1}{:l:}{Edgar} \\ \hdashline
 	\end{tabular}
   
  \caption{Output.}
  \label{fig:example-query-result}
  \end{minipage}
}
}}}
    \renewcommand{\thesubfigure}{e}
    \caption{Query plan in join-based \textcolor{nestedschemacolor}{$\blacksquare$} and flat RA \textcolor{flatschemacolor}{$\blacksquare$}.}
    \label{fig:query-plan-rete}
	\end{subfigure}
  \caption{Example property graph, textual query specification, query plans, and output of the query on the given graph. \\ The query finds persons \texttt{p} who are interested in some musical subject, but know at most two persons living in the same street.}
  \label{fig:running-example}
  \end{minipage}
\end{figure*}

\DecMargin{0.3em}
\begin{algorithm}[htb]
  \DontPrintSemicolon
  \KwData{$\mathit{op}$: NRA operator}
  \KwData{$\mathit{props}$: properties required by subsequent ops, initally $\emptyset$}
  \Fn{\InferRequiredProperties{op, props}}{
      $\mathit{props} \leftarrow \mathit{props} \union \ExtractProperties{op}$ \;
        \Switch{$\mathit{op}$}{
            \Case{is a nullary operator}{
              $\mathit{op.requiredProperties} \leftarrow \mathit{props}$ \;
            }
            \Case{is a unary operator}{
              \If {$op.type \in \{\projectionop, \groupingop\}$}{
                $\mathit{op.requiredProperties} \leftarrow \mathit{props}$ \;
              }
              $\mathit{op.child} \leftarrow   \InferRequiredProperties(\mathit{op.child}, \mathit{props})$ \;
            }
            \Case{is a binary operator}{
              $ \mathit{leftProps} \leftarrow \emptyset$;
              $ \mathit{rightProps} \leftarrow \emptyset$ \;
              \ForEach{$p \in \mathit{props}$} {
                 \uIf {$\text{vertex/edge of p} \in \mathit{op.left.nestedSchema} $}{
                   $\mathit{leftProps} \leftarrow \mathit{leftProps} \union \{p\}$
                 } \Else {
                   $\mathit{rightProps} \leftarrow \mathit{rightProps} \union \{p\}$
                 }
              }
              $\mathit{op.leftChild}  \leftarrow \InferRequiredProperties(\mathit{op.leftChild},  \mathit{leftProps}) $ \;
              $\mathit{op.rightChild} \leftarrow \InferRequiredProperties(\mathit{op.rightChild}, \mathit{rightProps}) $
            }
        }
    \Return $\mathit{op}$
  }
\caption{Infer required properties for NRA operators.}
\label{alg:schema-inferencing}
\end{algorithm}

\DecMargin{0.3em}
\begin{algorithm}[htb]
  \DontPrintSemicolon
  \KwData{$\mathit{op}$: NRA operator}
  \Fn{\ExtractProperties{op}}{
    \Switch{op}{
        \Case{is a $\{\projectionop, \groupingop\}$ operator}{
          $\mathit{ps} \leftarrow $ enumerate properties from $\textit{op}.\textit{projectionList}$ \;
        }
        \Case{is a $\selectionop$ operator}{
          $\mathit{ps} \leftarrow $ enumerate properties from $\textit{op}.\textit{condition}$ \;
        }
        \Case{is a $\sortandtopop$ operator}{
          $\mathit{ps} \leftarrow $ enumerate properties from $\textit{op}.\textit{orderAttributes}$ \;
        }
        \Case{is a $\thetaleftouterjoinop$ operator}{
          $\mathit{ps} \leftarrow $ enumerate properties from $\textit{op}.\textit{condition}$ \;
        }
        \Case{is an $\unwindop$ operator}{
          $\mathit{ps} \leftarrow $ enumerate properties from $\textit{op}.\textit{unwindAttribute}$ \;
        }
    }
    \Return $\mathit{ps}$
  }

\caption{Extract required properties from an NRA op.}
\label{alg:extract-properties}
\end{algorithm}

In \autoref{sec:gra}, we presented how to compile \opencypher queries to \gra, based on our previous work~\cite{DBLP:conf/adbis/MartonSV17}. However, the \gra representation poses two key challenges not sufficiently addressed in available IVM literature:
(1)~it uses graph-specific operators such as \emph{\expandtext} and \emph{\transitiveexpandtext}, and
(2)~it uses nested data structures.
To overcome these issues, we introduce two additional algebras: \emph{nested relational algebra} (\nra), which uses joins instead of expand operators, and \emph{flat relational algebra} (\fra), which uses flat relations instead of nested ones. We define a chain of steps which transform queries from \gra to \nra and from \nra to \fra (see the workflow in \autoref{fig:workflow}).

\subsection{Workflow example}

To demonstrate the workflow of our approach, we use the example graph in \autoref{fig:example-graph-2}, an extended and slightly altered version of the previous example graph in \autoref{fig:example-graph-1}.
The example query in~\autoref{fig:query-specification} finds \textsf{Persons} $p$ interested in some musical subject (including Music itself), who have less than 3 friends living in their street. \autoref{fig:query-plan-search} shows the \gra query plan for the example query. The first \lstinline{MATCH} clause of the graph query and a filtering condition is compiled to a sequence of a \emph{\getverticestext}, three \emph{\expandouttext}, and a \emph{\selectiontext} operator as shown in the bottom left branch of the tree. The transitive traversal on $\atom{SUBCLASS}$ edges is translated to a \emph{\transitiveexpandouttext}.
The pattern in the \lstinline{OPTIONAL MATCH} clause is compiled similarly, and combined with the other pattern using a \emph{\leftouterjointext}. Finally, the result is produced by a sequence of \emph{\groupingtext}, \emph{\selectiontext}, and \emph{\projectiontext} operators.

\subsection{Graph relational algebra to nested relational algebra}
\label{sec:gra-to-nra}

As a first transformation step, our workflow replaces \emph{\expandtext} operators with joins, resulting in an \nra query plan.

\fakeparagraph{One-hop expand}
We replace each \emph{\expandouttext} operator with a \emph{\jointext} on a \emph{\getedgestext} operator and similarly to the \emph{\expandintext} and \emph{\expandbothtext} operators, following the definitions in \autoref{sec:expand}.
Note that an \emph{\expandtext} operator following a \emph{\getverticestext} operator can be replaced with a single \emph{\getedgestext} operator, \eg:
\begin{align*}
&
\expandout{v}{w}{\labelsB}{e}{T}{1}{1}
\left(\getvertices{v}{\labelsA} \right)
\equiv
\left[
\getedgesdirected{v}{\labelsA}{w}{\labelsB}{e}{T}
\right],
\quad
\expandboth{v}{w}{\labelsB}{e}{T}{1}{1}
\left(\getvertices{v}{\labelsA} \right)
\equiv
\left[
\getedgesundirected{v}{\labelsA}{w}{\labelsB}{e}{T}
\right]
\\
&
\expandin{v}{w}{\labelsB}{e}{T}{1}{1}
\left(\getvertices{v}{\labelsA} \right)
\equiv
\projection{v, e, w}
\left[
\getedgesdirected{w}{\labelsB}{v}{\labelsA}{e}{T}
\right]
\end{align*}

\newcommand{\ger}{s}
\fakeparagraph{Transitive expand}
To map the \emph{\transitiveexpandtext} operator to joins, we introduce the \emph{\transitivejointext} operator~$r \transitivejoin{\atom{low}}{\atom{up}} s$. This operator joins relation~$r$ to the $k^\mathrm{th}$ selfjoin of relation~$s$ (where $\atom{low} \leq k \leq \atom{up}$), then returns the two endpoint vertices along with the intermediate edges. We only allow the right input of the transitive join operator to be a \emph{\getedgestext} operator. Therefore, with $\ger = \left[\getedgesdirected{v}{}{w}{}{e}{\types}\right]$, it can be defined as:
\begin{align*}
&
r \transitivejoin{\atom{low}}{\atom{up}} \ger
\equiv
r \join \Big(
\\
&
\quad \projection{v, \tuple{\tuple{1, x_1.\atom{edge}}, \ldots \tuple{\atom{low}, x_\atom{low}.\atom{edge}}}/E, w} 
\big( \ger_1 \join \ldots \join \ger_\atom{low} \big)
\union \\
&
\quad \projection{v, \tuple{\tuple{1, x_1.\atom{edge}}, \ldots \tuple{\atom{low}+1, x_{\atom{low}+1}.\atom{edge}}}/E, w} 
\big( \ger_1 \join \ldots \join \ger_{\atom{low}+1} \big)
\union\\
&
\quad \ldots \union \\[-1ex]
&
\quad \projection{v, \tuple{\tuple{1, x_1.\atom{edge}}, \ldots \tuple{\atom{up}, x_\atom{up}.\atom{edge}}}/E, w}
\big( \ger_1 \join \ldots \join \ger_\atom{up} \big) \Big),
\end{align*}
where $E$ is a nested attribute with schema $E\left(\atom{index}, \atom{edge}\right)$, similarly to the edge list attribute of the \emph{\transitiveexpandtext} operator (see \autoref{sec:expand}).
Using the \emph{\transitivejointext} operator, the \emph{\transitiveexpandtext} operators can be transformed as follows:
$$
\expandout{v}{w}{\labelsL}{E}{\types}{low}{up}
\left(r \right)
\equiv
r \transitivejoin{\atom{low}}{\atom{up}}
\left[
\getedgesdirected{v}{}{w}{}{E}{\types}
\right]
\join
\left( \getvertices{w}{\labelsL} \right)
$$

Note that the label constraint $\atom{\labelsL}$ is moved to a separate join on an additional \emph{\getverticestext} operator.
This is required as the label constraint does not have to be satisfied through all edges of $E$, only on its last vertex $w$. However, in most practical cases, transitive expand uses edge types which have the same vertex labels on their source and target vertices (\eg the $\atom{KNOWS}$ and $\atom{SUBCLASS\_OF}$ edge types of the example). In these cases, the label constraint can be kept during the traversal and the additional join can be omitted.\footnote{Complex transitive patterns can be generalised as regular path queries (RPQs), which have been studied in detail for one-time evaluation~\cite{DBLP:journals/siamcomp/MendelzonW95}, but not for incremental view maintenance. As of 2018, RPQs are supported to some extent by SPARQL~\cite{DBLP:journals/tods/PerezAG09} (in the form of \emph{property paths}) and PGQL~\cite{DBLP:conf/grades/RestHKMC16}.}
The expression in the example is translated as follows:
\begin{align*}
\expandout{fc}{c}{Class}{sos}{SUBCLASS\_OF}{0}{\infty} \left( r \right)
\equiv
r
\transitivejoin{0}{\infty}
\left[\getedgesdirected{fc}{}{c}{Class}{sos}{SUBCLASS\_OF}\right]
\end{align*}

\fakeparagraph{Example}
The \nra query plan of the example query is shown in \autoref{fig:query-plan-rete}, with the corresponding schema definitions in red~\textcolor{nestedschemacolor}{$\blacksquare$}.
The \emph{\expandtext} operators for the $\atom{KNOWS}$ / $\atom{INTEREST}$ edges and their child operators are combined to a \emph{\getedgestext} operator, while the rest of the \emph{\expandtext} operators are replaced with a left-deep tree of joins on \emph{\getedgestext} operators. Meanwhile, the \emph{\transitiveexpandouttext} operator is replaced with a \emph{\transitivejointext}. Other nodes of the \gra plan are left unchanged in the \nra plan.

\subsection{Nested relational algebra to flat relational algebra}
\label{sec:nra-to-fra}

Both \gra and \nra are nested algebras and represent vertex/edge properties as nested relations. As discussed in \autoref{sec:property-access}, we use a shorthand to access properties using a convenient syntax, \eg the projection operator in expression $\projection{\var{p}.\atom{name}}$ is allowed to use the value of the $\atom{name}$ property of vertex $\var{p}$.
However, due to the schema-free nature of property graphs, property keys of vertices/edges are not known in advance during compilation. The \gra and \nra formalisations work around this issue by treating the base relations of vertices and edges as nested (\nfii) relations. While this solves the problem in theory, it poses further challenges: nested relations are difficult to store efficiently and are not handled by most IVM algorithms. Hence, as the final step of the compilation, we transform the query plan to flat relational algebra (\fra).

\fakeparagraph{Schema inferencing}
We refer to the schema of \nra operators as the \emph{nested schema}, as it describes nested relations. In contrast, an \fra operator has a \emph{flat schema}, which contains all property keys required by the current operator and subsequent operators in the query plan.
The flat schema is determined by a two step \emph{schema inferencing} algorithm.

\begin{enumerate}
\item Starting from the root of the tree, we calculate \emph{required properties} for each operator, and push them down to the leafs. The corresponding pre-order traversal is described in \autoref{alg:schema-inferencing}, which relies on \autoref{alg:extract-properties} for extracting the properties from a given \nra operator.

\item Next, flat schemas of the \fra operators are calculated. For nullary operators, they are defined as a concatenation of the nested schema and the required properties; then, starting from nullary operators, the schema of each operator is calculated with a post-order traversal. Schemas are determined according to the conventions of relational algebra, except for
$\projectionop$ and $\groupingop$,
operators, where flat schemas are again defined as a concatenation of the nested schema and the required properties.
\end{enumerate}

\fakeparagraph{Example}
The \fra plan of the example query is shown in \autoref{fig:query-plan-rete}, with the corresponding schema definitions in blue~\textcolor{flatschemacolor}{$\blacksquare$}. Note that the required properties were added to the schema of each operator. For example, the \emph{\getedgestext} operator for $\atom{INTEREST}$ edges produces $\langle \var{p}, \var{pi}, \var{pt}, \uline{\var{p}.\atom{name}} \rangle$ quadruples, which include the property $\var{p}.\atom{name}$ used by operator $\projection{\var{p}.\atom{name}}$.

\section{View maintenance on Flat RA}
\label{sec:view-maintenance}

In \autoref{sec:nra-to-fra}, we defined steps to translate queries to an \fra query plan to allow evaluation with existing relational IVM algorithms such as \eg~\cite{DBLP:journals/ai/Forgy82,DBLP:journals/tkde/MirankerL91,DBLP:journals/tkde/Hanson96,DBLP:journals/tkde/HansonBC02,DBLP:conf/models/SzarnyasIRHBV14,DBLP:journals/scp/UjhelyiBHHIRSV15,PerPolQueryOptimization}. However, the rich set of operators required by \pg queries necessitates the combination of multiple techniques. In this section, we describe the IVM engine of our \emph{\ingraph} tool.

\subsection{Query evaluation in the Rete Network}

The query engine of \ingraph is built on the \emph{Rete algorithm}~\cite{DBLP:journals/ai/Forgy82,DBLP:conf/models/BergmannHRVBBO10,BergmannPhD,DBLP:journals/scp/UjhelyiBHHIRSV15}, which was originally developed to incrementally handle production rules in rule-based expert systems.
Unlike \emph{algebraic} IVM techniques (\eg~\cite{DBLP:journals/tkde/QianW91,DBLP:conf/sigmod/GriffinL95}), which derive delta queries to maintain the results of the target query, the Rete algorithm follows a \emph{procedural} approach that maintains each relational algebraic operator separately. This makes it more composable and simpler to extend.

In essence, the Rete algorithm employs a \emph{space-time tradeoff}~\cite{DBLP:conf/sigmod/RossSS96} to speed-up query processing evaluation. First, it builds a \emph{propagation network}, which follows the topology of the flat relational algebra query plan. Each operator is subscribed to the output of its child operators and propagates it result to its parent operator. Calculations start from the leaf nodes which correspond to nullary operators \emph{\getverticestext} and \emph{\getedgestext}.
IVM in the Rete network is achieved by extensive caching: nodes in the Rete network store interim results which allows efficient computation for small updates.

\fakeparagraph{Example}
In the example query of \autoref{fig:running-example}, the $\left[ \getedgesopdirected \right]$~operator for $\atom{INTEREST}$ subscribes to the indexer and receives
$
\{
\langle a, 1, n, \atom{"Alice"} \rangle, \ldots,
\langle e, 5, v, \atom{"Edgar"} \rangle
\}
$
tuples. Other \emph{\getedgestext} operators are populated similarly, and the results are propagated through the unary and binary relational algebraic operators, producing the initial query result (\autoref{fig:example-query-result}).

\subsection{Cache maintenance in the Rete Network}
\label{sec:cache}
Changes in the data, including the initial load phase, are represented logically as changes in nullary operators $\left( \getverticesop \right)$, $\left[ \getedgesopdirected \right]$, and $\left[ \getedgesopundirected \right]$. Changes are propagated through the actor network as \emph{update messages} containing positive and negative change sets (representing insertions and deletions, respectively). For each unary and binary \fra operator, incremental maintenance operations are defined for both insertions and deletes.

\fakeparagraph{Example}
In \autoref{fig:running-example}, $\atom{Person}$ ``Edgar'' gains interest in ``Rock'' music. This change is represented as adding a tuple $\langle e, 6, r, \atom{``Edgar"} \rangle$ to the $\left[ \getedgesopdirected \right]$ operator for type $\atom{INTEREST}$, which is propagated through the network, adding a new tuple $\langle \atom{``Edgar"} \rangle$ to the result set (\autoref{fig:example-query-result}).

\subsection{Data representation and indexing}
\label{sec:indexer}
The \emph{\ingraph} prototype is a memory-only engine with no permanent storage. To allow efficient lookup of vertices, edges, and their properties, it uses an indexer layer. The indexer is capable of performing lookups based on ids/labels, and sending \emph{notifications} on updates of the data. In general, lookups are cheaper when more constraints are provided, \eg it is cheaper to get the set of edges when both the edge type and the source/target labels are specified compared to when only the edge type is known.
This is the key reason why our approach uses compound operators (such as \emph{\getedgestext} which takes one type and two label constraints), instead of using primitive operators as building blocks.

In the relational operators of the execution engine, tuples correspond to the \emph{flat schema}. This ensures that the internal data representation of operators is compact %
and allows each operator to perform its computation based on local data without turning to the indexer, thus satisfying the actor model.

\subsection{Programming model}

The implementation of \ingraph uses the \emph{actor programming model}, which captures concurrent computations
as \emph{actors} (with isolated mutable states)
that communicate by \emph{asynchronous immutable messages}.
Once a query is compiled, the engine builds an \emph{actor network} based on the Rete network, \ie it instantiates one actor for each operator. Nullary operators in the query plan are captured as subscriptions to the indexer, which is responsible to perform efficient lookups and generate change notifications. As actors have no shared state, they can be run in parallel and even distributedly. We previously demonstrated this with the \mbox{\textsc{IncQuery-D}} engine that implemented distributed IVM on top of RDF graphs~\cite{DBLP:conf/models/SzarnyasIRHBV14}.

\begin{table*}[!htb]
  \setlength\extrarowheight{2.5pt}
  \setlength\tabcolsep{3.6pt}
  \begin{tabular}{@{}rllcccccc@{}}
  	\toprule
  	                                  \bf ref. & \bf venue        & \bf contributions                                                                                        & \bf A/P & \bf bag & \bf \nfii & \bf null & \bf aggr. & \bf ord. \\ \midrule
  	      \cite{DBLP:conf/sigmod/BlakeleyLT86} & SIGMOD'86        & determining irrelevant updates, maintenance of select--project--join views                                                 &   A+P   &   \no   &   \no   &   \no    &    \no    &   \no    \\
  	         \cite{DBLP:journals/tkde/QianW91} & TKDE'91          & change propagation equations for relational alg.; fixed in~\cite{DBLP:journals/tkde/GriffinLT97}         &    A    &   \no   &   \no   &   \no    &    \no    &   \no    \\
  	         \cite{DBLP:conf/sigmod/GuptaMS93} & SIGMOD'93        & counting algorithm (non-recursive views), DRed algorithm (recursive views)                                         &    P    &  \yes   &   \no   &   \no    &    \no    &   \no    \\
  	       \cite{DBLP:conf/sigmod/ColbyGLMT96} & SIGMOD'96        & change propagation equations for bag alg., incl. aggregation but no group-by                             &    A    &  \yes   &   \no   &   \no    &  \maybe   &   \no    \\
  	      \cite{DBLP:conf/dbpl/KawaguchiLMR97} & DBPL'97          & extending IVM techniques to maintain views defined over a nested data model                              &    A    &  \yes   &  \yes   &   \no    &    \no    &   \no    \\
  	          \cite{DBLP:conf/dbpl/LibkinW97a} & DBPL'97          & maintaining the transitive closure of directed graphs using a SQL-like language                          &    A    &  \yes   &  \yes   &   \no    &   \yes    &   \no    \\
  	        \cite{DBLP:conf/sigmod/MumickQM97} & SIGMOD'97        & group-by-aggregation, summary-deltas for representing changes                                            &    P    &  \yes   &   \no   &   \no    &   \yes    &   \no    \\
  	     \cite{DBLP:journals/tkde/GriffinLT97} & TKDE'97          & improved change propagation equations for relational algebra                                             &    A    &   \no   &   \no   &   \no    &    \no    &   \no    \\
  	    \cite{DBLP:journals/sigmod/GriffinK98} & SIGM. R.'98 & change propagation equations for semijoins, antijoins and outer joins                                    &    A    &   \no   &   \no   &   \yes   &    \no    &   \no    \\
  	            \cite{DBLP:conf/ideas/LiuVM99} & IDEAS'99         & incremental equations for the operators of the nested model                                              &    A    &   \no   &  \yes   &   \no    &    \no    &   \no    \\
  	       \cite{DBLP:conf/vldb/PalpanasSCP02} & VLDB'02          & maintenance of non-distributive aggregate functions                                                      &    P    &  \yes   &   \no   &   \yes   &   \yes    &   \no    \\
  	         \cite{DBLP:conf/er/DimitrovaER03} & ER'03            & order-preserving maintenance of XQuery views                                                             &    A    &  \yes   &  \yes   &   \no    &  \maybe   &   \yes   \\
  	          \cite{DBLP:journals/is/GuptaM06} & IS'06            & generalised summary-deltas, group-by-aggregations, outer joins; fixed in~\cite{DBLP:conf/icde/LarsonZ07} &    A    &  \yes   &   \no   &   \yes   &   \yes    &   \no    \\
  	            \cite{DBLP:conf/icde/YiYYXC03} & ICDE'03          & top-$k$ views                                                                                            &    P    &   \no   &   \no   &   \no    &    \no    &   \no    \\
  	           \cite{DBLP:conf/icde/LarsonZ07} & ICDE'07          & outer joins and aggregation                                                                              &    P    &  \yes   &   \no   &   \yes   &   \yes    &   \no    \\
  	    \cite{DBLP:journals/vldb/KochAKNNLS14} & VLDBJ'14         & higher-order IVM, viewlet transformations, the DBToaster system                                          &    A    &  \yes   &   \no   &  \maybe  &  \maybe   &  \maybe  \\ \bottomrule

  \end{tabular}
	\caption{%
    Overview of related literature on IVM techniques, presented in order of appearance. Notation: \yes~fully supported, \maybe~supported to some extent, \no~not supported, A/P: algebraic/procedural.  }
  \vspace{-1.5ex}
	\label{table:related-work}
\end{table*}

\section{Related work}
\label{sec:related-work}

\subsection{Incremental view maintenance algorithms}

Covering the rich set of features required by property graph queries -- ranging from expressing negative structural conditions to unnesting and reachability queries -- requires different IVM algorithms.
Surveys on IVM approaches were presented in paper~\cite{DBLP:journals/debu/GuptaM95},
book~\cite{Gupta:1999:MVT:310709}, 
and monograph~\cite{DBLP:journals/ftdb/ChirkovaY12}. However, even such comprehensive surveys did not cover challenges
\ref{challenge:schema-optional}--\ref{challenge:meta} and \ref{challenge:reachability}--\ref{challenge:higher-order} presented in \autoref{sec:introduction}.

A preliminary work on algebraic view maintenace was presented in~\cite{DBLP:journals/tkde/QianW91}.
Its algorithm was improved in~\cite{DBLP:journals/tkde/GriffinLT97}, 
co-authored by Griffin and Libkin who produced one of the seminal papers in the field~\cite{DBLP:conf/sigmod/GriffinL95}.
Later studies add extensions to support additional operators: aggregations~\cite{DBLP:conf/views/Quass96}, 
semijoins/outer joins~\cite{DBLP:journals/sigmod/GriffinK98}, 
order-preserving maintenance~\cite{DBLP:conf/er/DimitrovaER03}, 
and outer joins/aggregations~\cite{DBLP:conf/icde/LarsonZ07}.
Techniques for IVM on object-oriented data were presented in~\cite{DBLP:journals/kais/LiuVM03}.
\autoref{table:related-work} shows an overview of IVM techniques and their applicability to bags, \nfii data, null values, complex aggregations and ordering, along with their categorisation to algebraic/procedural.\footnote{Note that the distinction between algebraic/procedural techniques is not always clear, \eg the approach of~\cite{DBLP:conf/sigmod/BlakeleyLT86} is considered as algebraic in some works~\cite{DBLP:journals/ftdb/ChirkovaY12} and procedural in others~\cite{DBLP:conf/er/DimitrovaER03}.}

Due to the rise of interest in efficient graph processing techniques, recent efforts aimed to design \emph{relational join algorithms} that were specifically suited to handle \emph{subgraph matching} efficiently~\cite{DBLP:journals/jacm/NgoPRR18}. Incrementalized join algorithms suited for subgraph matching were published in~\cite{DBLP:journals/pvldb/AmmarMSJ18,DBLP:journals/pvldb/IdrisUVVL18}.

\subsection{Rule-based expert systems}

IVM has been used extensively in the context of \emph{rule-based expert systems} (also known as \emph{production systems}), supported by \emph{discrimination networks}. Notable approaches include Rete~\cite{DBLP:journals/ai/Forgy82}, TREAT~\cite{DBLP:journals/tkde/MirankerL91}, and Gator~\cite{DBLP:journals/tkde/HansonBC02}.
In expert systems, users formulate \emph{rules} (or \emph{productions}), which have a \emph{left-hand side} (LHS) and a \emph{right-hand side} (RHS). As described in~\cite{DBLP:journals/tkde/MirankerL91}, a \emph{rule engine} (or \emph{production system interpreter}) repeatedly executes a cycle of three operations: (1)~match, (2)~conflict set resolution, and (3)~act.

A performance comparison of the Rete and TREAT algorithms is given in~\cite{DBLP:conf/icde/WangH92} and~\cite{DBLP:conf/vldb/BrantGLM91}. ``An algebraic approach to rule analysis in expert database systems'' was presented in~\cite{DBLP:conf/vldb/BaralisW94}.
A heavily modified version of the Rete algorithm is used in the Drools~\cite{DBLP:conf/agtive/Proctor11} rule-based expert system.

\subsection{Query languages}

Paper~\cite{DBLP:journals/csur/AnglesABHRV17} contains a detailed survey on modern graph query languages. It discusses popular data models, defines two categories of query functionalities (graph patterns and navigational expressions) and presents important concepts such as \emph{matching semantics}. According to this categorisation, our work focuses on \emph{graph patterns}. In the following, we discuss query languages for graph pattern matching and implementation that provide (some degree of) incremental view maintenance.

\fakeparagraph{Cypher and openCypher}
Early attemps to formalise the Cypher language were presented in~\cite{DBLP:conf/edbt/HolschG16,DBLP:conf/adbis/MartonSV17}, which use graph relational algebra to capture the semantics of the language. The formal semantics for Cypher's core were presented in~\cite{DBLP:conf/sigmod/FrancisGGLLMPRS18}.

\emph{Implementations.}
Graphflow~\cite{DBLP:conf/sigmod/KankanamgeSMCS17} is an active openCypher database, which bears the closest similarity to our approach. Its language extends Cypher with user-defined functions that trigger on new matches, but it lacks support for advanced language features such as negative/optional edges and reachability.

\fakeparagraph{SPARQL}
Of existing graph query languages, SPARQL is the best understood in terms of semantics and complexity~\cite{DBLP:journals/tods/PerezAG09}. In the last decade, multiple works targeted IVM for SPARQL.

\emph{Implementations.}
Diamond~\cite{Diamond12} uses the Rete algorithm to evaluate SPARQL queries on \emph{distributed RDF data}. During the evaluation of a query, it identifies additional tuples by dereferencing URLs, turning to remote servers and feeding new data elements to the Rete network.
\mbox{INSTANS}~\cite{DBLP:conf/semweb/Rinne12} uses the Rete algorithm to perform \emph{complex event processing} on streaming RDF data.
Strider~\cite{DBLP:journals/pvldb/RenCKLBRZK17} is a recent research prototype supporting continuous SPARQL queries.

\fakeparagraph{VIATRA Query Language}
Graph pattern matching has been used extensively in the domain of model-driven engineering, \eg by the \viatra framework. Its \viatra Query Language (VQL) is based on Datalog~\cite{DBLP:conf/icmt/BergmannURV11}, and supports recursive queries, subpattern calls, along with some aggregations.

\emph{Implementations.}
The \viatra framework uses the Rete algorithm to perform efficient model validation and transformation operations over graph models~\cite{DBLP:journals/scp/UjhelyiBHHIRSV15}. 
\iqd~\cite{DBLP:conf/models/SzarnyasIRHBV14} is a \emph{distributed} incremental graph query engine, which uses a query language based on VQL and operates on RDF graphs.

\section{Conclusion and future work}
\label{sec:summary}

In this paper, we presented an approach towards incrementally querying property graphs. Our approach compiles graph queries to relational graph algebra, then translates them to nested relational algebra and finally converts them to flat relational algebra. The resulting expression is then maintained using relational IVM techniques. 

Up to our best knowledge, this is the first work dedicated to study \emph{incremental view maintenance on property graphs}. As such, we believe it opens up interesting research directions:

\begin{itemize}
\item It allows using recent advancements in incremental join algorithms such as~\cite{DBLP:journals/pvldb/AmmarMSJ18} and~\cite{DBLP:journals/pvldb/IdrisUVVL18} for PG queries.
\item It facilitates the development of cost-based optimisation techniques for property graph queries~\cite{DBLP:phd/dnb/Gubichev15,DBLP:conf/icmt/VarroD13}.
\item The presented incremental evaluation techniques can be used to define \emph{graph views} on top of RDBMSs~\cite{DBLP:conf/grades/Xirogiannopoulos17}.
\item It can be extended by adapting algorithms designed to perform graph-specific operations, \eg \emph{graph search}~\citeprog{DBLP:journals/sosym/VarroDWS15}, and \emph{impact analysis techniques}~\cite{DBLP:conf/models/RederE12}.

\end{itemize}

\section*{Acknowledgements}

This work was partially supported by NSERC RGPIN-04573-16 and the MTA-BME Lend\"ulet Cyber-Physical Systems Research Group.
The authors would like to thank G\'abor Bergmann, M\'arton B\'ur, B\'alint Hegyi, J\'ozsef Makai, Krist\'of Marussy, M\'arton Elekes, Oszk\'ar Semer\'ath, D\'avid Szak\'allas, Muhammad Idris, and Semih Salihoglu, along with the Graphflow team, the DBToaster team and the Cypher Language Group for discussions.

\bibliographystyle{ACM-Reference-Format}
\bibliography{ms}

\end{document}